%% file: main.tex
\documentclass[journal=jpcafh,manuscript=article]{achemso}
\usepackage{longtable}
\usepackage{multirow}
\usepackage{amsmath}

\usepackage[usenames,dvipsnames]{color}
\usepackage{soul}
\usepackage{bm}

\usepackage{amssymb}
\usepackage{microtype}
\usepackage[normalem]{ulem}  

\usepackage{subcaption}
\usepackage{tikz}
\usetikzlibrary{arrows.meta} 
\usepackage[outline]{contour} 
\contourlength{1.4pt}
\pgfdeclarelayer{background}
\pgfsetlayers{background,main}
\usetikzlibrary{matrix,chains,positioning,decorations.pathreplacing,arrows,shadows}
\usetikzlibrary{fadings}
\tikzset{>=latex} 
\usepackage{xcolor}
\colorlet{myred}{red!80!black}
\colorlet{myblue}{blue!80!black}
\colorlet{mygreen}{green!60!black}
\colorlet{mydarkred}{red!30!black}
\colorlet{mydarkblue}{blue!40!black}
\colorlet{mydarkgreen}{green!30!black}
\tikzstyle{node}=[thick,circle,draw=myblue,minimum size=22,inner sep=0.5,outer sep=0.6]
\tikzstyle{node in}=[node,green!20!black,draw=mygreen!30!black,fill=mygreen!25]
\tikzstyle{node hidden}=[node,blue!20!black,draw=myblue!30!black,fill=myblue!20]
\tikzstyle{node out}=[node,red!20!black,draw=myred!30!black,fill=myred!20]
\tikzstyle{connect}=[thick,mydarkblue] 
\tikzstyle{connect arrow}=[-{Latex[length=4,width=3.5]},thick,mydarkblue,shorten <=0.5,shorten >=1]
\tikzset{ 
  node 1/.style={node in},
  node 2/.style={node hidden},
  node 3/.style={node out},
}

\usepackage{multicol} \usepackage{wrapfig} \usepackage{enumitem}
\usepackage{bm} \usepackage{outlines} 
\SectionNumbersOn

\author{Silvan K\"aser} \affiliation[University of Basel]{Department of
  Chemistry, University of Basel, Klingelbergstrasse 80, CH-4056
  Basel, Switzerland.}
  \author{Luis Itza Vazquez-Salazar} \affiliation[University of Basel]{Department of
  Chemistry, University of Basel, Klingelbergstrasse 80, CH-4056
  Basel, Switzerland.}
\author{Markus Meuwly} \affiliation[University of Basel]{Department of Chemistry, University
  of Basel, Klingelbergstrasse 80, CH-4056 Basel, Switzerland.}\email{m.meuwly@unibas.ch}
\author{Kai T\"opfer} \affiliation[University of Basel]{Department of
  Chemistry, University of Basel, Klingelbergstrasse 80, CH-4056
  Basel, Switzerland.}  \email{kai.toepfer@unibas.ch}

\title{Neural Network Potentials for Chemistry: Concepts, Applications and Prospects}

\begin{document}

\tableofcontents
\clearpage

\begin{abstract}
\noindent
Artificial Neural Networks (NN) are already heavily involved in
methods and applications for frequent tasks in the field of
computational chemistry such as representation of potential energy
surfaces (PES) and spectroscopic predictions. This perspective
provides an overview of the foundations of neural network-based
full-dimensional potential energy surfaces, their architectures,
underlying concepts, their representation and applications to chemical
systems. Methods for data generation and training procedures for PES
construction are discussed and means for error assessment and
refinement through transfer learning are presented. A selection of
recent results illustrates the latest improvements regarding accuracy
of PES representations and system size limitations in dynamics
simulations, but also NN application enabling direct prediction of
physical results without dynamics simulations. The aim is to provide
an overview for the current state-of-the-art NN approaches in
computational chemistry and also to point out the current challenges
in enhancing reliability and applicability of NN methods on larger
scale.
\end{abstract}

\section{Introduction}

\noindent
The \textit{in silico} modeling of chemical and biological processes
at a molecular level is of central importance in today's research and
will be crucial for future challenges of mankind.\cite{Zhou2019big}
The modeling often requires a trade-off between accuracy and
computational cost: Quantum chemical calculations (e.g. \textit{ab
  initio} molecular dynamics), at a high level of theory, can be very
accurate but also come at a high computational cost rendering the
approach impractical except for rather small molecules. Empirical
force fields, on the other hand, provide a computationally
advantageous approach that scales well with system size but the
possibility to carry out quantitative studies is limited due to the
assumptions underlying their formulation. Thus, computationally
efficient and accurate modelling techniques are required for
quantitative molecular simulations.\cite{friederich2021machine}
\\

\noindent
In this regard, Machine Learning (ML) techniques have emerged as a
powerful tool to satisfy such demands for force field models which are
limited, in principle, by the accuracy of \textit{ab initio} methods
and allow an efficiency approaching that of empirical force
fields.\cite{unke2021machine} Motivated by the advances in
computational chemistry techniques and the continuous growth of the
performance of computer hardware (Moore's
law\cite{moore1998cramming}), ML is becoming a daily tool for modeling
molecules and materials. By definition, ML methods are data-driven
algorithms based on statistical learning theory with the aim of
generating numerical methods that generalize to new data, not used in
the learning process.\cite{vapnik1999nature,meuwly2021machine} This
capability renders ML methods highly appealing for modelling molecular
systems. It even reaches levels where some authors believe that the
use of ML techniques will constitute the "fourth paradigm of
science",\cite{agrawal2016perspective} bridging the gap from
atomic-scale molecular properties towards macroscopic properties of
materials\cite{bartok2017machine,hirn:2020} and one of the drivers for
a revolution of the simulation techniques of
matter.\cite{aspuru2018matter} The enthusiasm is reflected in the
appearance of an extensive number of ML models and their application
in computational chemistry.  \\

\noindent
Some of the most important publications have focused on the study of
potential energy surfaces (PESs), which contain all the information
about the many-body interactions of a molecular system including
stable and metastable structures.\cite{noe2020machine} At the same
time, it is possible to extract a considerable amount of information
from PESs including the atomic forces driving the dynamics of
molecular systems, reactions and structural transitions, and atomic
vibrations.\cite{behler2021perspective} Additionally, it has been
proposed that the chemical information contained in a chemical bond,
therefore in the PES, can help in the exploration of chemical
space.\cite{shaik2013one} In a recent work,\cite{vazquezsalazar2021}
it was found that the exploration of chemical space can be improved by
adding adequate information from the configurational space represented
by the PES.\\

\noindent
Over the past several decades several ML-based methods have been 
used to represent continuous PESs.\cite{behler2022review,unke2021machine,bowman:2018,pronobis2020kernel}
While a number of those are briefly mentioned below, the focus of 
the present work is on NN-based approaches.
Kernel-based methods provide an efficient solution to highly non-linear
optimization problems\cite{pronobis2020kernel} by finding a
\textit{representation} of the problem which encodes the distribution
of the data in a complete, unique and efficient
way.\cite{lilienfeld2016baml} There is a large number of possible
representations of chemical space that can be used in kernel
methods. Examples include Coulomb Matrices\cite{rupp2012fast}, Bag of
Bonds (BoB)\cite{hansen2015machine}, Histograms of Distance, Angles
and Dihedrals (HDAD)\cite{faber2017prediction}, Spectrum of London and
Axilrod-Teller-Muto (SLATM)\cite{lilienfeld2020slatm},
Faber-Christensen-Huang-von Lilienfeld (FCHL)\cite{lilienfeld2020fchl}
and Smooth Overlap of Atomic Positions (SOAP).\cite{bartok2013} A
comprehensive review of representations for kernel and non-kernel
methods can be found in Ref.~\citenum{musil2021physics}. It should be
noted that variations of kernel methods, such as for Gaussian
processes\cite{deringer2021gaussian} which assume a
Bayesian/probabilistic point of view for the solution of the problem
or the reproducing kernel Hilbert space (RKHS)
method\cite{ho1996general,unke2017toolkit} which uses polynomials as
support functions have been extensively discussed in the literature.
While the remainder of the perspective is mainly dedicated to NN-based
approaches, many alternative interpolation and representation methods
for PES construction 
exist. These include,
e.g.  modified Shepard interpolation\cite{collins2002molecular},
(interpolative) moving
least-squares\cite{lancaster1981surfaces,farwig1986multivariate,bender2014potential},
permutationally invariant polynomial (PIP) PESs by least-squares
fitting\cite{bowman2009pip}, or least absolute shrinkage and selection
operator (LASSO) constrained least-squares.\cite{mizukami2014compact}
Several of these approaches have been recently described, reviewed and
compared.\cite{unke2021machine,dawes2018construction,bowman2022pip}
\\

\noindent
NNs are inspired by the biological model of the intricate networks
formed by the brain and how information is
passed.\cite{mcculloch1943logical} The ideas underlying NNs date back
to 1960 when "the perceptron" was presented by
Rosenblatt.\cite{rosenblatt1958perceptron} However, computational and
theoretical limitations inhibited the development of
NNs.\cite{minsky1969perceptron,o2022undercover} It was not until 1970
with the development of the automatic differentiation and the
introduction of backpropagation\cite{rumelhart1986learning} that NN
models continued to develop. Still, large scale applications were rare
until the beginning of the 21st century when considerably more
powerful computer hardware became available. In chemistry, the
application of NN models dates back to 1990s with first applications
in analytical and medicinal chemistry.
\cite{gasteiger1993nn,zupan1999neural} Regarding PES representation,
the first application of NNs can be tracked back to the same
decade.\cite{sumpter1992potential,blank1995neural} Nowadays, NNs are
the most common ones from the field of ML models for the use in
chemistry-related applications that are focused on the generation and
study of PESs.  Some examples of popular NN-based schemes for PES
fitting include the High Dimensional Neural Network (HDNN)
method\cite{behler2007gnn,ko2021fourth}, Deep Tensor Neural Network
(DTNN)\cite{schuett2017dtnn}, SchNet\cite{schutt2018schnet},
ANI\cite{smith2017ani}, or PhysNet\cite{unke2019physnet}, among
others.\\

\noindent
The purpose of the present perspective is to provide a birds-eye view
and an outlook into the conception, generation and use of NN based
PESs for the exploration of chemical systems.  Additionally, we will
present some of the current challenges in the development and
application of NN models for the study of PESs.  The remainder of the
present work is structured as follows. A brief introduction to the
theoretical background of PESs and NNs is provided in Section
2. Section 3 discusses existing NN architectures with emphasis on
structural information and current developments in the field.  Section
4 describes the construction of a PES from the initial sampling to the
validation and refinement of the generated models and Section 5
discusses knowledge transfer that allows obtaining PESs at high levels
of theory with less data. Selected applications for chemical systems
showcasing the concepts introduced and including NN models in
established atomistic dynamics models are described in Section
6. Applications of NN models that skip dynamics simulation to predict
physical observables are shown in Section 7. Section 8 describes some
of the current challenges that we consider critical for the
development and enhancement of the current models and the field in
general, followed by a short conclusion.\\

\section{Theoretical Background}
This section introduces the concept of PESs, the principles underlying
NNs, their building blocks, such as dense layers and activation
functions. A more in-depth overview of descriptors for chemical
structures and representative examples of frequently used neural
network potentials (NNPs) is given in the next section. In terms of
nomenclature, italic symbols denote scalars or functions and bold
symbols are $n-$dimensional tensors ($n\geq1$) with the special case
of a one-dimensional spatial vector (e.g. position or distance)
denoted as italic symbol with vector arrow.\\

\subsection{Potential Energy Surfaces}
\begin{figure}
    \centering \includegraphics[scale=0.75]{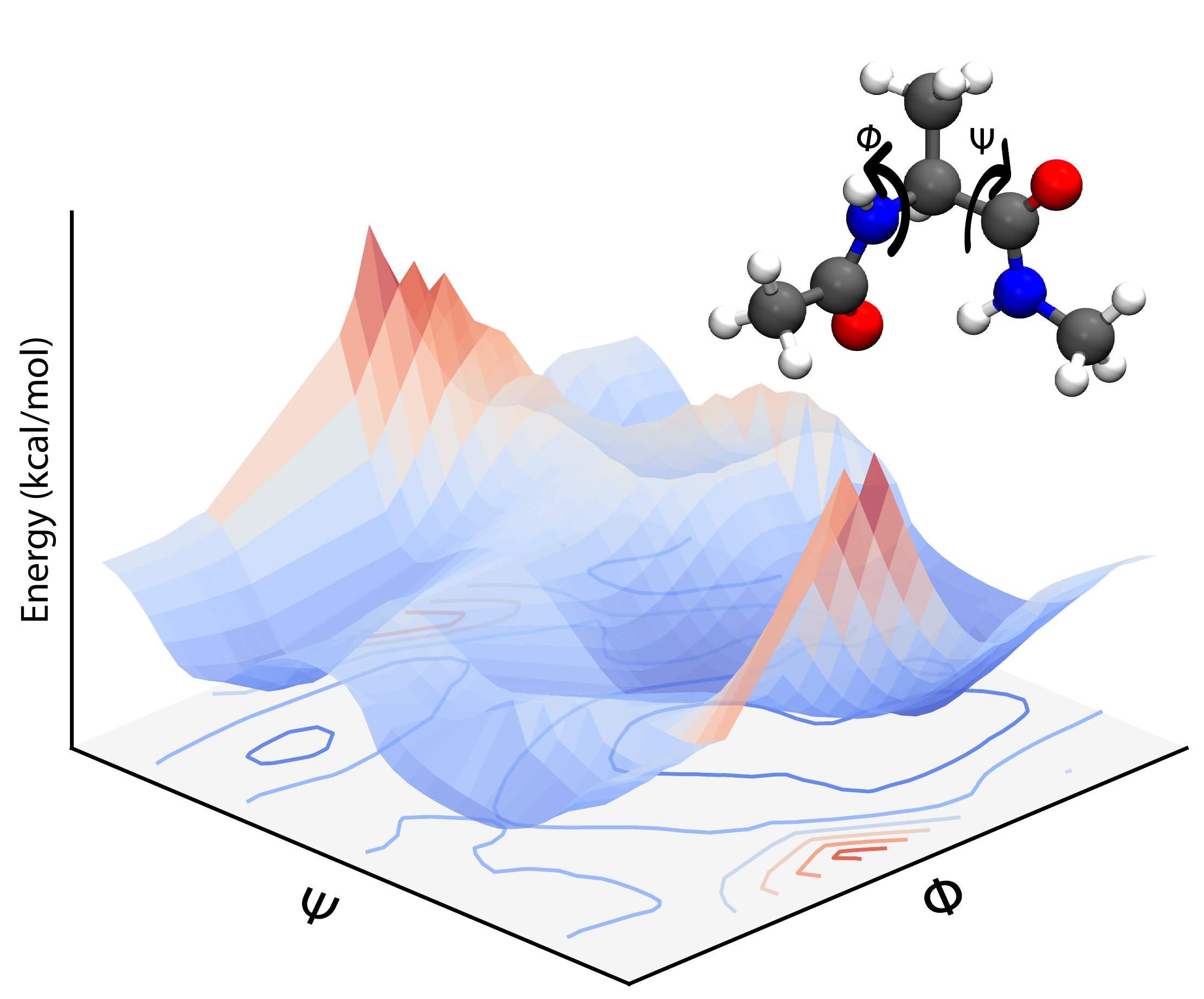}
    \caption{A two-dimensional PES for the dialanine molecule
      calculated at the MP2 level with the 6-31G** basis set along
      dihedral angles $\Phi$ and $\Psi$. A representation of the
      molecule (ball and stick) indicating the dihedral angles ($\Phi,
      \Psi$) calculated is given as well. The bottom gives the
      projection of the 2D PES.}
    \label{fig:PES}
\end{figure}

The energetics of a molecular system can be described by solving the
electronic Schr\"odinger Equation (SE). Unfortunately, the SE can only
be solved exactly for simple, single-electron atomic systems. In order
to obtain solutions for many-electron systems, it is necessary to
introduce approximations. The Born-Oppenheimer approximation
(BOA)\cite{born1927quantentheorie}, also called \textit{the most
  important approximation in quantum
  chemistry},\cite{tannor2007introduction} assumes that the coupling
between the nuclear and electronic motion can be neglected because the
mass of the nuclei is several orders of magnitude larger than the mass
of the electrons. Under this assumption, it is possible to rewrite the
total wavefunction $\Psi$, which is a solution of the SE, as the
product of a nuclear wavefunction $\chi(\mathbf{R})$ with nuclear
positions $\mathbf{R}$ and the electronic wavefunction
$\psi(\mathbf{r} ; \mathbf{R})$ with electron coordinates $\mathbf{r}$
for a fixed configuration of nuclear positions
\begin{equation}
    \Psi(\mathbf{r},\mathbf{R}) =
    \psi(\mathbf{r};\mathbf{R})\cdot\chi(\mathbf{R}).
    \label{eq:bo}
\end{equation}
As a consequence, the electronic wavefunction can be obtained by
solving the electronic time-independent SE:
\begin{align}
\hat{H}_{\rm e}\psi_\lambda(\mathbf{r};\mathbf{R}) =
\left[\hat{T}_{\rm e} + \hat{V}_{\rm ne} + \hat{V}_{\rm ee}\right]
\psi_\lambda(\mathbf{r};\mathbf{R})
= \epsilon_\lambda(\mathbf{R})\psi_\lambda(\mathbf{r};\mathbf{R})
\label{eq:etise}
\end{align}
Here, $\hat{H}_{\rm e}$ is the electronic (spin-free) Hamiltonian
describing the kinetic energy of the electrons $\hat{T}_{\rm e}$,
the Coulomb interaction between the nuclear and electron charges
$\hat{V}_{\rm ne}$ and the electron-electron interaction $\hat{V}_{\rm
  ee}$. The solution is the electronic wavefunction $\psi_\lambda$ and
electronic energy $\epsilon_\lambda$ for the electronic state
$\lambda$. The so-called adiabatic PES of an atomic system
$E^{\rm BO}_\lambda(\mathbf{R})$ in electronic state $\lambda$ constitutes
an effective potential for the nuclear dynamics. It is obtained by the
sum of the Coulomb repulsion $V_{\rm nn}$ between the nuclei with
nuclear charge $Z_i$ for the total number of atoms $N$, and the
respective electronic energy at the associated nuclear
positions.\cite{jensen2017introduction}
\begin{equation}
    E^{\rm BO}_\lambda(\mathbf{R}) = V_{\rm nn}(\mathbf{R}) +
    \epsilon_\lambda(\mathbf{R})
    \label{eq:bo-pes}
\end{equation}
Equation \ref{eq:bo-pes} defines a PES as a $(3N-6)-$dimensional
function that can be approximated as an analytical function which is,
however, a challenging task. Often, one can only report
low-dimensional cuts of such high-dimensional hypersurfaces and one
example is shown in Figure~\ref{fig:PES}.  Alternatively, equation
\ref{eq:bo-pes} suggests that there should be a mapping between the
total electronic energy of a molecular system and the combination of
position of the nuclei and the set of nuclear charges
$\{Z_i\}_{i=1}^N$. This is the starting point for a ML-based approach
described in the following.\\

\noindent
PESs lie at the heart of computational
chemistry.\cite{lewars2011computational} From the relationship between
structure and potential energy $E$, it is possible to derive many
molecular properties by taking derivatives with respect to a
perturbation such as atomic positions $\mathbf{R}$, an external
electric $\Vec{\mathcal{E}}$ or magnetic field $\Vec{\mathcal{B}}$,
which require additional coupling terms in the Hamiltonian and an
analytical representation of the PES.\cite{jensen2017introduction}
Following this, a general response property takes the form
\begin{equation}
    {\rm Property} \propto \dfrac{\partial^{(n+m+l)}E}{\partial
      \mathbf{R}^n \partial \Vec{\mathcal{E}}^m \partial
      \Vec{\mathcal{B}}^l }
    \label{eq:pes_derv}
\end{equation}
where $n,m,l$ indicate the order of the derivative with respect to the
perturbation. Derivatives of Equation~\ref{eq:pes_derv} provide, e.g.,
the forces $\mathbf{F}=-\partial E /\partial \mathbf{R}$ that
constitute the foundation of MD simulations and structure optimization
schemes. The second derivatives $\partial^{2} E / \partial
\mathbf{R}^{2}$ gives access to the Hessian matrix from which the
harmonic frequencies of molecular vibrations can be obtained.  Other
properties such as the dipole moment ($\Vec{\mu}=-\partial E /
\partial \Vec{\mathcal{E}}$) or the molecular polarizability
($\Vec{\alpha}=-\partial^{2} E / \partial \Vec{\mathcal{E}}^{2}$) are
directly related to experimental observables such as the Infrared (IR)
or Raman spectra.\cite{keith2021combining} Mixed derivatives also
provide IR absorption intensities ($\partial^{2} E / \partial
\Vec{\mathcal{E}} \partial \mathbf{R}$) or the optical rotation in
circular dichroism ($\partial^{2} E / \partial \Vec{\mathcal{E}}
\partial \Vec{\mathcal{B}}$).\\

\noindent
Given the versatility and usefulness of PESs, a wealth of approaches
to construct PESs have been designed over the years and new ML schemes
are proposed with high frequency. Especially NNs have been shown to be
general function
approximators\cite{gybenko1989approximation,hornik1991approximation}
by the universal approximation theorem\cite{hornik1989multilayer} and
hence seem particularly useful to learn intricate relationships such
as the PES or even external perturbations.\\

\subsection{Artificial Neural Networks}
Artificial NNs (NNs, henceforth) represent a family of computer
algorithms and form a subgroup of ML. Nowadays, NNs are applied in
diverse areas including, among others, health
care\cite{shailaja2018machine}, medical
imaging\cite{litjens2017survey}, self-driving
cars\cite{grigorescu2020survey}, high-energy
physics\cite{guest2018deep}, particle physics and
cosmology\cite{carleo2019machine},
genetics\cite{angermueller2016deep}, chemical
discovery\cite{von2020retrospective}, reaction
planning.\cite{wei2016neural,segler2018planning}\\

\begin{figure}
\centering
\begin{subfigure}{\textwidth}
\centering
	\input{tikz_figs/nn.tex}
\end{subfigure}
\hfill

\begin{subfigure}{0.4\textwidth}
    \input{tikz_figs/neuron.tex}
\end{subfigure}
\hspace{60pt}
\vspace{-60pt}
\begin{subfigure}{0.4\textwidth}
\centering
\includegraphics[scale=0.4]{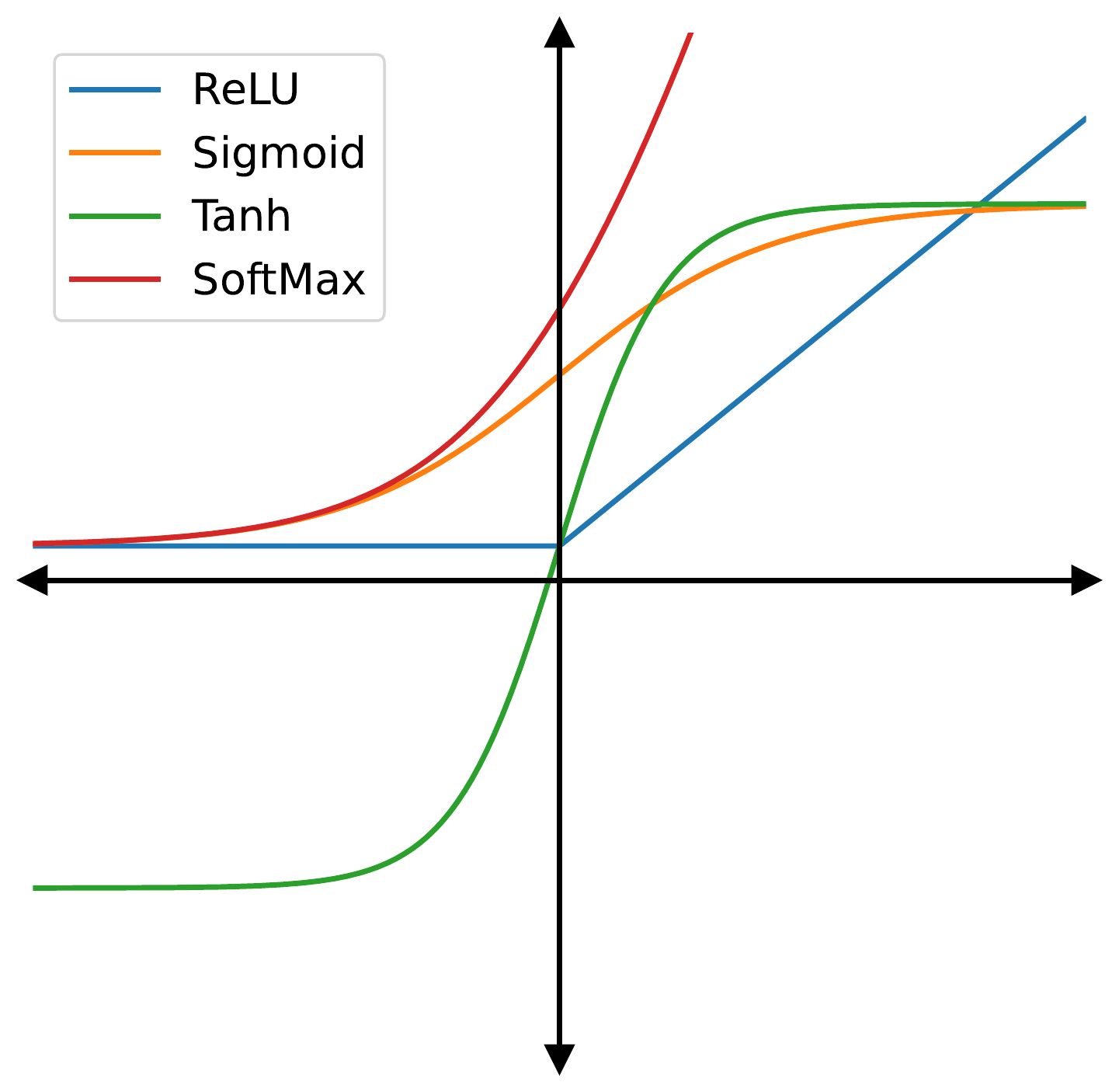}
\end{subfigure}
\vspace{60pt}
\caption{Neural network and its building blocks. (A) Schematic of a NN
  model with an input layer (green), $N$ hidden layers (blue) and an
  output layer (red). (B) Illustration of a node inside the hidden
  layers.  Bottom right (C): Examples of common activation functions.}
\label{fig:NN_draw}
\end{figure}

\noindent
Typically, a NN consists of an input layer, a predefined number of
hidden layers and an output layer (see
Figure~\ref{fig:NN_draw}A). Deep NNs comprise a larger number of
hidden layers while a NN with only one or two hidden layers is a
shallow NN. Each layer contains a defined number of nodes (or neurons)
that connect to the nodes of the following layer and each connection
is associated with weights and biases.  \\

\noindent
The elementary units of NNs are so-called dense layers, which linearly
transform an input vector $\mathbf{x}$ to an output vector
$\mathbf{y}$ according to
\begin{align}
    \mathbf{y} = \mathbf{Wx} + \mathbf{b}. 
\end{align}

\noindent
Here, $\mathbf{W}= \{ w_{ij} \}_{i,j=1}^{N,M}$ and $\mathbf{b}=\{ b_i
\}_{i=1}^{N}$ are the weights (a matrix) and biases (a
vector),\cite{unke2021machine} $M$ is the dimension of the input and
$N$ the number of nodes. The combination of a dense layer with a
nonlinear activation function (Figure~\ref{fig:NN_draw}B) transforms
the input $\mathbf{x}$ to an output $\mathbf{y}$ that serves as
"input" to the following (hidden) layer.
\begin{equation}
    \mathbf{h}_{i} = \sigma\left(\mathbf{W}_i\mathbf{x} + \mathbf{b}_i\right) 
    \label{eq:hidden_layer}
\end{equation}
Modelling non-linear relationships requires the combination of at
least two dense layers with an activation function $\sigma$ according
to
\begin{align}
    \mathbf{y} = \mathbf{W}_{i+1}\sigma\left(\mathbf{W}_i\mathbf{x} +
    \mathbf{b}_i\right) + \mathbf{b}_{i+1} =
    \mathbf{W}_{i+1}\mathbf{h}_{i} + \mathbf{b}_{i+1}
\end{align}
While such shallow architectures are in principle capable of modelling
any functional relationship, deeper variants thereof are usually
preferred due to improved performance and
parameter-efficiency.\cite{eldan2016power,cohen2016expressive,telgarsky2016benefits,lu2017expressive}
The functional form of the NN is characterized by the number of layers
$L$ and number of nodes $N$ in a given layer. With increasing $L$ and
$N$ the functional form becomes more flexible, however, overfitting
requires careful attention since the obtained form has no underlying
physical meaning.\cite{behler2017first} A fully connected deep NN is
given by the following relation
\begin{align}
\mathbf{y}=\mathbf{W}_{i_{L}}^{L}\sigma(\mathbf{W}_{i_{L}i_{L-1}}^{L-1}\sigma(\cdots
\sigma(\mathbf{W}_{i_{2}i_{1}}^{1}(\sigma\mathbf{W}_{i_{1}i_{0}}^{0}\mathbf{x}+\mathbf{b}_{i_{0}}^{0})+
\mathbf{b}_{i_{1}}^{1})\cdots)+
\mathbf{b}_{i_{L-1}}^{L-1})+\mathbf{b}_{i_{L}}^{L}
\end{align}
which is usually followed by a linear transformation in the final
output layer to yield the prediction $\mathbf{y}_{L+1}$. If the NN is
used to construct a PES, a chemical descriptor $\mathbf{x}$ is mapped
onto one or multiple scalar values $\mathbf{y}=\{V\}$, which are the
energies of one or several electronic states for an atomic
configuration.  \\

\noindent
As mentioned above, the flexibility and power of a NN is related to
the number of layers and nodes but the ability to obtain highly
non-linear relationships between inputs and outputs is a consequence
of the use of appropriate activation functions (Figure
\ref{fig:NN_draw}C).  Activation functions usually satisfy particular
mathematical properties, including differentiability (crucial for
computing forces or vibrational frequencies)\cite{schutt2020learning}
and \textit{smoothness}, that simplifies the optimization of the model
and increasing the quality of the prediction of energy and
forces.\cite{montavon2020introduction}\\

\noindent
Besides the architecture of a NN, the actual training (or
``learning'') step is important, too. Training comprises the parameter
fitting process of the weights and biases to match the prediction
$\mathbf{y}(\mathbf{x})$ to the reference results $\mathbf{t}$ for a
set of $N_\mathrm{data}$ data points. The accuracy of the fit is
measured by monitoring a loss function $\mathcal{L}$ which has the
general form\cite{montavon2020introduction}:
\begin{equation}
    \mathcal{L} = \dfrac{1}{N_\mathrm{data}}
    \sum_{n=1}^{N_\mathrm{data}}[\mathbf{y}(\mathbf{x_n})-\mathbf{t_n}]^{m}
    + \omega
    \label{eq:loss}
\end{equation}
The value of $m$ in Equation \ref{eq:loss} mostly takes the value
$m=1$ or $2$ ($L_{1}$ or $L_{2}$ norm) and $\omega$ can be a
regularization term that helps to improve the generalizability of the
model and to prevent overfitting (i.e. the model is fitted perfectly
against training data losing generalizability). Different loss
functions for fitting NNs can be used as
well.\cite{vazquez2022uncertainty} In general, the loss function is
highly nonlinear and is minimized iteratively by a gradient descent
algorithm which, preferably, can find the best solution despite
potentially many local minima.\cite{keith2021combining} For PES
fitting, convergence behaviour and accuracy can be improved by
including additional information such as atomic forces or dipole
moments (or other properties of the system) in the loss function.\\

\section{Neural Networks for Potential Energy Surfaces}
The use of NNs to represent PESs of molecular systems started in the
1990s. However, initially it was only possible to include a few
degrees of freedom.
\cite{gasteiger1993nn,doren1995nn,clary1996internal,scheffler2004nnpes,doren2005nnpes}
Applicability and transferability of NNs to larger systems and with
different system compositions were improved by the approach proposed
by Behler and Parrinello who decomposed the total energy of a system
into atomic contributions\cite{behler2007gnn}
\begin{equation}
    E = \sum_{i=1}^{N}E_{i} ~~~.
    \label{eq:e_decom}
\end{equation}
Here, $N$ is the total number of atoms and $E_{i}$ is the energy of
atom $i$ that can be predicted by one or multiple NNs (e.g. one for
each atomic element). The inputs are local, atom-centered descriptors
that encode the local chemical environment around atom $i$. Rooted in
Equation \ref{eq:e_decom}, the so-called high-dimensional NNP
(HDNNP),\cite{behler2007gnn,behler2011acsf} was introduced and
followed by further
models.\cite{smith2017ani,e2018dpmd,bin2019eann,kitchin2020singlenn,
  schuett2017dtnn,schuett2017schnet,barros2018hipnn,unke2019physnet,isayev2019aimnet,isayev2021aimnetme,unke2021spookynet,behler2011hdnnp3,yao2018tensormol,goedecker2015cent,small2020bpopnn,ko2021fourth,schuett2021painn,satorras2021egnn,batzner2022nequip}
It is important to note that most of the commonly used models are
based on the decomposition of the energy in atomic contributions,
although models that represent the energy as the sum of bond energies
have also been
proposed.\cite{behler2012pairhdnnp,parkhill2017bimnn,sherrill2020apnet}
In the following, we will focus on NNs that decompose the potential
energy into atomic contributions.\\

\subsection{Descriptors}
\label{sec:descriptors}
All NNs are based on a local representation of the chemical
environment to correctly predict the reference
data.\cite{bartok2013,knoll2015descr,huo2022unified,uhrin2021descr,rupp2022descr}
Such representations require descriptors that, most importantly, are
(i) \emph{invariant} with respect to transformations including
translation, rotation and permutation of same elements, (ii)
\emph{unique} by showing changes when transformation that modify the
predicted property are applied and (iii) \emph{continuous} and
\emph{differentiable} with respect to the atomic coordinates to
determine forces for molecular
simulations.\cite{huo2022unified,rupp2022descr} Based on the type of
local representation that incorporates all the conditions above, NNPs
can be classified into two major categories: those with predefined and
those with learnable descriptors.\cite{behler2022review}\\

\subsubsection{Predefined Descriptors}
Encoding the atomic environment by descriptors that fulfill the
previously described characteristics has been a challenge since the
early beginnings of the development of NN models and it is still an
area of active development. Some of the requirements for a 'good'
descriptor can be matched with simple transformations of the Cartesian
atom positions. For example, rotational and translational invariance
can be obtained by using internal
coordinates.\cite{noe2020machine,behler2021four} However,
permutational invariance is more difficult to incorporate. A solution
to this problem is the use of PIPs\cite{bowman2009pip} as input for a
NNP, which are still extensively used for small molecule
PESs.\cite{jiang2013permutation,li2013permutation,jiang2016potential,shao2016communication,fu2018ab,chen2020fitting}
Other solutions are based on using symmetrized input coordinates or
symmetry incorporated in the NN.\cite{behler2021four}\\

\noindent
A better solution to the problems described above was found with
predefined descriptors introduced by Behler and Parrinello in 2007
with the development of the HDNNP.\cite{behler2007gnn,behler2011acsf,smith2017ani,kitchin2020singlenn,
  behler2011hdnnp3,goedecker2015cent,small2020bpopnn,ko2021fourth}
These descriptors, termed atom-centered symmetry functions
(ACSF)\cite{behler2011acsf,behler2015constructing} or
variations\cite{smith2017ani,yao2018tensormol} thereof are the
prevalent predefined descriptors for NNPs in the literature.\\

\noindent
Originally, the local chemical environment of atom $i$ is encoded by
sets of radial- and angular-type symmetry functions $G^\mathrm{rad}_i$
and $G^\mathrm{ang}_i$ for each element or element combination of
atoms $j$ and $k$ individually. A modified version of Gastegger and
coworkers, on the other hand, combines them linearly with a weighting
factor depending on the respective atoms' element number $Z_j$ and
$Z_k$.\cite{gastegger2018wacsf}
\begin{equation}
   G^\mathrm{rad}_i (\eta, R_s) =
        ~\sum_{j \neq i}^{N} 
        g(Z_j) \cdot e^{-\eta(R_{ij} - R_s)^2} \cdot f_c(R_{ij})
        \label{eq:bp_rad}
\end{equation}
\begin{equation}
    G^\mathrm{ang}_i (\zeta, \lambda, \eta, R_s) =
    \begin{aligned}[t]
        & ~2^{1 - \zeta}
        \sum_{j \neq i}^{N} 
        \sum_{k \neq i,j}^{N} 
        h(Z_j,Z_k) \cdot \left( 1 + \lambda \cos{\theta_{ijk}} \right)^\zeta &\\
        & \cdot e^{-\eta(R_{ij} - R_s)^2} 
        \cdot e^{-\eta(R_{ik} - R_s)^2} 
        \cdot e^{-\eta(R_{jk} - R_s)^2} &\\
        & \cdot f_c(R_{ij}) \cdot f_c(R_{ik}) \cdot f_c(R_{jk}) ~.
    \end{aligned}
\label{eq:bp_ang}
\end{equation}
In this version of weighted ACSF (wACSF) $R_{ij}$, $R_{ik}$, $R_{jk}$
are pair distances and the angle $\theta_{ijk}$ is defined between the
vectors $\vec{R}_{ij}$ and $\vec{R}_{ik}$.  The contributions to the
symmetry function are limited by the cutoff function $f_c(R)$ which
monotonically decrease from 1 to 0 at the cutoff separation $R_c$. The
parameter $\lambda \in \{-1,1\}$ determines the maxima of the cosine
term at $\theta_{ijk} = 0^\circ$ or $180^\circ$.  The resolution and
size of the descriptor are determined by the choice and number of
combinations of hyperparameters $\eta$ and $R_s$ for the radial
symmetry functions $G^\mathrm{rad}_i$ as well as $\zeta$ and $\eta$
for the angular symmetry functions $G^\mathrm{ang}_i$. The functions
$g(Z_{j})$ and $h(Z_j,Z_k)$ are the element-dependent weighting
functions for which even simple expressions such as $g(Z_j) = Z_j$ and
$h(Z_j, Z_k) = Z_j Z_k$ yielded satisfactory results.\cite{gastegger2018wacsf}
\\

\noindent
Regarding the ACSF representation, each descriptor is a vector for
which the length depends on combinations of the sizes of respective
hyperparameters $\eta$, $R_s$ and $\zeta$ with size $N_\mathrm{par}$
but also the number of different chemical elements $N_\mathrm{el}$ in
the atomic system.  These are $N_\mathrm{par} \cdot N_\mathrm{el}$ for
radial-type and $N_\mathrm{par} \cdot N_\mathrm{el}(N_\mathrm{el} +
1)/2$ for angular-type symmetry functions. The size of the radial- and
angular-type wACSF simply scales by the respective combination of the
hyperparameters. HDNNPs with descriptor sizes of 32 wACSFs, 220 ACSFs
and 35 ACSFs were trained using the energies of the molecules in the
QM9 database with up to five elements. The mean absolute error of the
validation and test set is reported even lower for the model with
wACSFs (1.84 and 1.83\,kcal/mol, respectively) than the 220 ACSFs
(2.49 and 2.39\,kcal/mol) and 35 ACSFs (7.57 and
7.40\,kcal/mol).\cite{gastegger2018wacsf} \\

\noindent
ACSFs commonly apply expensive trigonometric cutoff functions but
computationally much cheaper polynomial cutoff functions can be
designed for the same functionality.\cite{dellago2019caching} Further
improvement in the performance is achieved by replacing the
exponential function and cosine in radial- and angular-type symmetry
function with dedicated polynomials with essentially no loss in
accuracy.\cite{dellago2021} The speedup is shown by MD simulations of
360 water molecules using a HDNNP that performs about 1.8 times faster
with polynomial symmetry and cutoff functions than with the original
ACSFs.\cite{dellago2021} \\

\noindent
Another type of fixed descriptors was introduced by E and coworkers in
their Deep Potential (DP) model.\cite{e2018dpmd,wang2018deepmd} These
are based on the construction of a local coordinate frame which
assures the required invariances. Once the positions of the atoms are
transformed by a translation and rotational matrix, the local
coordinates can be used to construct the descriptor based on radial
and/or angular information. However, this descriptor cannot ensure
smoothness because of the uncertainty in the choice of the local frame
that can lead to discontinuities.\cite{wen2022deep} E and coworkers
proposed the Deep Potential-Smooth Edition (DP-SE)
model\cite{zhang2018end} to solve the mentioned issue by enforcing
continuity of the descriptor by multiplying the local coordinate
system with a continuous and differentiable function and modifying the
embedding matrix to recover two-body and three-body terms of the
descriptor.\cite{wen2022deep} \\

\noindent
In addition to the ACSF functions and the DP descriptor, there are
other descriptors that utilize the concept of neighbourhood density
functions.\cite{khorshidi2016amp,unke2018reactive} For this type of
descriptors the information about the local environment of atom $i$ up
to a cutoff radius is represented by a density function
$\rho(\mathbf{R_i})$ depending on the nuclear charge $Z_j$ and
position $\mathbf{R_j}$ of neighbouring atoms $j$.
\begin{equation}
    \rho(\mathbf{R}) = \sum_{j,||\mathbf{R}_{j}||\leq R_c}
    Z_{j}\delta(||\mathbf{R}-\mathbf{R}_{j}||)
    \label{eq:ndf}
\end{equation}
Here, $\delta$ is the Dirac delta function. In order to use this
function in a NNP, it is necessary to expand $\rho(\mathbf{R_i})$ in a
basis set of fixed dimension. For Gaussian-type basis functions, the
ACSF functions are obtained.\cite{khorshidi2016amp} Other interesting
expansions include the use of Zernike basis sets in which radial basis
functions and spherical harmonics polynomials are
used.\cite{unke2018reactive} \\

\noindent
A major problem of using predefined descriptors is that it requires a
certain degree of knowledge to define the hyperparameters
appropriately.\cite{behler2011acsf,smith2017ani,kitchin2020singlenn,
  behler2011hdnnp3,goedecker2015cent,small2020bpopnn,ko2021fourth}
Even though some of the hyperparameters can be optimized during the
training as well,\cite{yao2018tensormol,bin2019eann} a poor choice of
hyperparameters can lead to limited resolution of certain atomic
displacements with quasi-constant descriptors and degenerate values of
the predicted energy for different geometrical
structures.\cite{ceriotti2020incomplete,goedecker2022fingerprints} The
disadvantages of fixed descriptors motivated the emergence of NNPs
which directly learn a suitable representation of atomic positions and
element types.\cite{schutt2020learning,unke2021machine}\\

\subsubsection{Learnable Descriptors}
\begin{figure}
\centering
  \input{tikz_figs/mpnn.tex}
\caption{Message-passing principle visualized on a chain of three
  nodes with initial feature vectors $h_i^{t=0}$ representing the
  colour fraction red, green, blue on the mixed colour of node
  $i$. The message operation $M_t$ corresponds to the addition of the
  feature vectors within in cutoff range and the update operation
  $U_t$ corresponds to an addition of $h_i^{t+1} = h_i^{t} + m_i^{t}$
  and scaling that $sum \{ h_i^{t+1} \} = 1$. Although it is outside
  the cutoff radius $R_c$, after two iterations the feature vector of
  node 1 ($h_1^{t=2}$) contains a fraction (information) of the
  initial feature vector from node 3 (visualized by the blue coloured
  path).}
\label{fig:mpnn}
\end{figure}
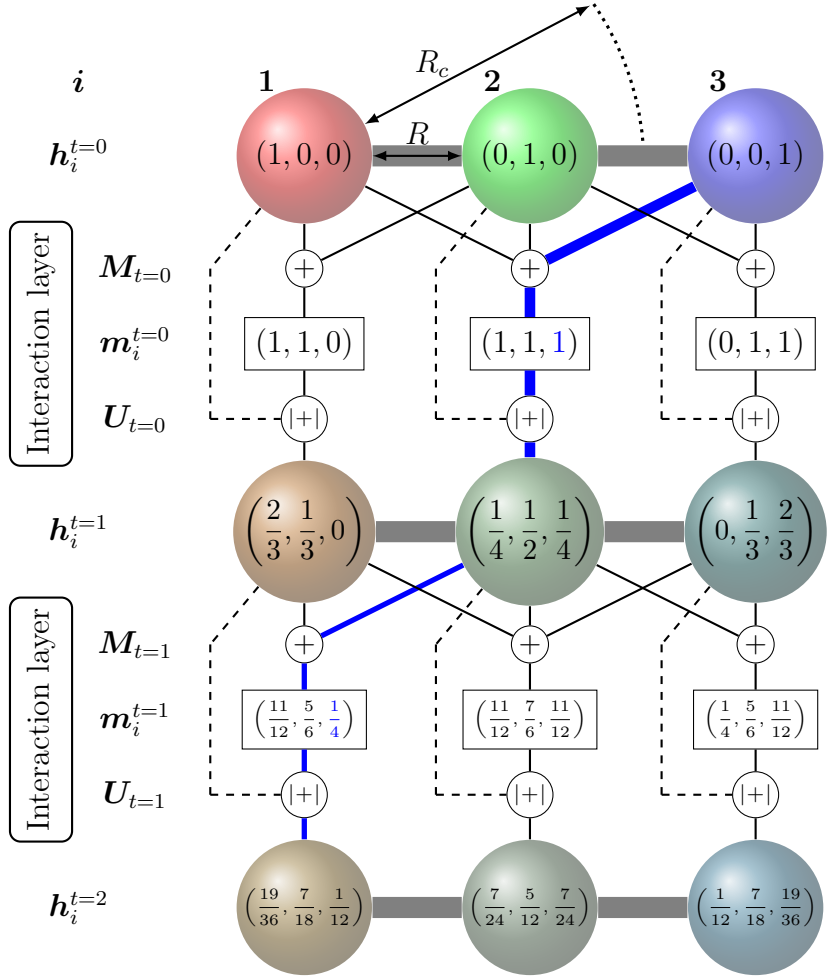

\noindent
The concept of learnable descriptors originates from graph neural
networks.\cite{scarselli2009gnn} In general, atoms are regarded as
nodes (not to be confused with nodes of NN layers), each associated
with a feature vector, which are connected to their neighbouring atoms
within a cutoff sphere by so-called edges. Information between the
nodes is passed along the edges over multiple iterations to encode the
necessary chemical interaction.\\

\noindent
The feature vectors of each node with length $N_f$ are randomly
initialized as a function of the atoms' nuclear charge, that is
iteratively updated by a message vector encrypting structural
information and feature vectors of the atoms within a cutoff sphere by
passing through \textit{interaction layers} which ensure the required
invariances. Figure \ref{fig:mpnn} visualizes the message passing
principle on a linear chain of nodes (atoms) with distance $R$, where
the feature vector $h_i^{t}$ at each iteration step $t$ corresponds to
the ratio of the colours red, green and blue to the mixed colour. In
each interaction layer, the feature vectors of node $i$ and connected
nodes within cutoff range $R_c$ are combined by a message function
$M_t$ (addition) to the message vector $m_i^{t}$. Note that this
message function does not encode distances $R$. The message vector
$m_i^{t}$ is combined with the feature vector $h_i^{t}$ by an update
function $U_t$ (addition and scaling to linear sum of 1) to form a
refined feature vector $h_i^{t+1}$ that contains information of the
surrounding nodes. Message and update functions usually include the
transformation of feature with update vectors by a NN.  For an
iteration step $t > 1$, this approach allows that information from
nodes that are outside of the cutoff range can still be incorporated
in a feature vector of a given node $i$ indirectly. This means that
for the case illustrated in Figure \ref{fig:mpnn}, the feature vector
$h_1^{t=2}$ of node 1 contains a fraction of blue colour after two
iterations ($\frac{1}{12}$) that is passed from node 3 via node 2.\\

\noindent
Many of the more recently developed NNPs
\cite{schuett2017dtnn,schuett2017schnet,barros2018hipnn,
  isayev2019aimnet,unke2019physnet,unke2021spookynet,isayev2021aimnetme,
  schuett2021painn,satorras2021egnn,batzner2022nequip} apply such
atom-wise feature vector approaches and are called message-passing NNs
(MPNNs).\cite{duvenaud2015mpnn,gilmer2017mpnn} Depending on the MPNN
model, the atomic feature vectors of either the final iteration or
each iteration are passed to a specific NN and transformed into the
desired quantity (e.g. energy).\\

\noindent
Feature vectors with higher number of elements $N_f$ and more complex
message and update functions including bond distance and direction
dependencies allow higher resolution of the structural encoding.  In
common NNPs, the number of elements in the feature vectors $N_f$ range
from about 64 to 128 per element. A larger number might increase the
risk of overfitting.\cite{barros2018hipnn} Similarly, a larger number
of message passing iterations improves the representation of the
structural features but the potential energy accuracy usually shows
sufficient saturation after three iteration
($t=3$).\cite{schuett2017dtnn,schuett2017schnet,barros2018hipnn,isayev2019aimnet,schuett2021painn}\\

\subsection{Architectures} 
\label{subsec:arch}
Given that the field of NNPs is very active, it is impossible to
describe all the available NN architectures. Hence this section is not
a comprehensive review of all possible architectures but rather a more
history-guided view of architectures and what functionalities were
included in subsequent development steps.\\

\noindent
Initial models use NNs as a method for the fitting of PES only (no
forces).\cite{carrington2021rev} These models were limited to small
molecules in gas phase and were fitted to energies of \textit{ab
  initio} calculations via a many-body
expansion\cite{malshe2009development} or a high-dimensional model
representation.\cite{manzhos2006random} Therefore, these models take
energies and positions to predict coefficients for a defined
functional form. These models already achieved spectroscopic accuracy
for small molecules.\cite{behler2021perspective} \\

\noindent
The introduction of the HDNNP with the concept of decomposing the
molecular energy into atomic contributions (Equation \ref{eq:e_decom})
changes the paradigm of NNPs. A new challenge was encoding the local
environment information sufficiently well for an accurate energy
prediction that lead to the two main approaches of predefined or
learnable descriptors. The main development of NN architectures with
predefined descriptors goes towards more sophisticated descriptors to
encode atom-centered properties which are then provided to standard
fully-connected feed-forward NNs.\cite{hellstrom2020high} NN
architectures with learnable descriptors and the MPNN approach differ
in their message and update functions within an interactions layer.
\\

\noindent
The first MPNN proposed was the deep tensor neural network
(DTNN)\cite{schuett2017dtnn} by Sch\"utt and coworkers that had been
further improved into the, to this day, popular SchNet
model.\cite{schuett2017schnet} An interaction layer in SchNet includes
so called continuous-filter convolutional layers that have already
been used in image or sound processing.\cite{schuett2017schnet} A
combination of the popular predefined ACSF descriptors and learnable
ones was proposed by Isayev and coworkers and their atoms-in-molecule
NN model (AIMNet).\cite{isayev2019aimnet} Modified ACSF descriptors
from the ANI architecture were used for initialization of atomic
structure feature vectors, combined with atomic information feature
vectors and passed through the interaction layer.  \\

\noindent
Although these models already achieve good accuracy, long range
interactions between chemical compounds can only contribute to the
total energy if the information is included in or passed to the
descriptor by a sufficiently long cutoff range $R_c$. Systems with
strong electrostatic interactions, especially with highly polar or
ionic chemical species, requires larger cutoffs but at the cost of
higher computational demand.\cite{unke2019physnet} One solution is to
add a Coulomb term to the atomic energy contributions which includes
electrostatic interactions between atomic charges $q$ predicted by the
NN model.
\begin{equation}
  E = \sum_{i=1}^{N} \left[ E_{i} + \frac{1}{2}
    \sum_{j>i}^{N}\dfrac{q_{i}q_{j}}{R_{ij}} \right]
  \label{eq:e_decom_pcharges}
\end{equation}
The earliest NN model using Equation~\ref{eq:e_decom_pcharges} was
introduced by Artrith and Behler in 2011 that trains a separate NN
with reference charges from a Hirshfeld population
analysis.\cite{behler2011hdnnp3} Another approach is applied by the
TensorMol model that predict atom charges by fitting the \emph{ab
  initio} and physically determinable molecular dipole moment to the
predicted one computed by the atom charges.\cite{yao2018tensormol} \\

\noindent
Additional physically motivated interactions, such as dispersion
interactions, were also included in the TensorMol model but have been
employed in PhysNet, too. PhysNet is based on the MPNN architecture
and was developed by Unke and Meuwly.\cite{unke2019physnet} It does
not only add an energy contribution from the DFT-D3 dispersion
correction scheme\cite{grimme2011dftd3} but also modifies
Equation~\ref{eq:e_decom_pcharges} by applying a damping function that
smoothly damps Coulomb interactions for small atom distances to avoid
singularities
\begin{equation}
  E = \sum_{i=1}^{N} \left[ E_{i} + \frac{1}{2}
    \sum_{j>i}^{N}q_{i}q_{j}\cdot\chi(R_{ij}) \right] + E_\mathrm{D3}~.
  \label{eq:e_decom_physnet}
\end{equation}
$E_\mathrm{D3}$ is the DFT-D3 dispersion correction and the damping
function $\chi(\mathbf{R}_{ij})$ is defined as:
\begin{equation}
    \chi(R_{ij}) = \phi(2 R_{ij}) \dfrac{1}{\sqrt{R_{ij}^{2} + 1}} +
    (1 - \phi(2 R_{ij}))\dfrac{1}{R_{ij}}~.
\end{equation}
A continuous behaviour is ensured by the cutoff function
$\phi(R_{ij})$.  \\

\noindent
Although adding a Coulomb term to NNPs improves the description of
long range interactions while the atomic charges still depend on the
local chemical environment.\cite{behler2021four} However, chemical
systems are inherently non-local. Therefore, the approximation breaks
down for systems with changes in the total charge state
(i.e. ionization, protonation or deprotonation), electronic
delocalization or spin density rearrangements.\cite{unke2021spookynet}
These effects are difficult to capture with NN architectures which
model changes in the atom charges by local perturbations.  \\

\noindent
The most recent generation of NNPs addresses the problem of non-local
charge transfer by using different strategies.  The first work
dedicated to the issue of charge equilibration was the charge
equilibration via NN technique (CENT) developed by Ghasemi and
coworkers.\cite{goedecker2015cent} The CENT algorithm equilibrates the
charge density to minimize the electrostatic energy which depends on
environment-dependent atomic electronegativity and hardness besides
the charge-charge interaction.  Inspired by CENT, Behler and coworkers
introduced their fourth generation HDNNP (4G-HDNNP) model where NNs
are trained to predict environment-dependent atomic
electronegativities (constant element-specific hardness) and the
charge equilibration yields the reference atomic
charges.\cite{ko2021fourth} In a second training step, NNs provided
with ACSFs and the atomic charge information are trained to predict
the short-range atomic energy contributions which sum up with the
electrostatics to the correct reference energy and forces.  \\

\noindent
SpookyNet is a MPNN model and introduced by Unke and coworkers that
treats the problem of non-locality by creating an embedding for
charges and spin.\cite{unke2021spookynet} It is capable to predict
molecular systems with different spins and charged states as provided
in the reference data set within one single model.  The general idea
of predicting PESs of chemical systems for different electronic states
and their coupling strength within one model is an area of active
research.\cite{westermayr2020machine} One model in this direction that
can be mentioned is SchNarc\cite{marquetand2020schnarc} that combines
the SchNet model with the surface hopping including arbitrary
couplings (SHARC)\cite{Mai2018WCMS} code.  \\

\noindent
So far, we have been reporting the effort to improve the models
accuracy by introducing more physically motivated
interactions. However, current developments for MPNNs focus on passing
spatial directions between atoms to the NN that allow the prediction
of atom-centered tensorial properties such as atomic
polarizability.\cite{gasteiger2019directional,miller2020relevance,schuett2021painn}
Providing solely distance information inherently ensures translational
and rotational invariance for atom-centered scalar properties
(predictions do not change with respect to, e.g., rotation of the
molecule).  The challenge with directional information is rotational
equivariance which means that predicted atom-centered directional
properties $\vec{f}(\mathbf{R})$ keep its amplitude but change in
direction equivalent to a rotation $\mathcal{A}$ of the molecular
coordinates $\mathbf{R}$.
\begin{equation}
  \mathcal{A} \cdot \vec{f}(\mathbf{R}) = \vec{f}(\mathcal{A} \cdot \mathbf{R})
  \label{eq:enn}
\end{equation}
MPNN that encode directional information (directional message passing) and
fulfill Equation~\ref{eq:enn} are called equivariant NNs (ENNs).
\cite{thomas2018tensor,smidt2021euclidean}
\\

\noindent
ENNs have been proven to be data-efficient and capable of providing
better predictions of tensorial quantities (i.e. dipole, quadrupole
moments) than invariant models.  ENN models with different
modifications were suggested to include directional information and
assure equivariance.  Some of them are PaiNN,\cite{schuett2021painn}
NeuqIP,\cite{batzner2022nequip} and
NewtonNet.\cite{haghighatlari2022newtonnet} Still one of the best
performing ENNs on the QM9 data set is DimeNet, where rotational
equivariance is achieved by representing the local chemical
environment of an atom by spherical 2D Fourier-Bessel basis with
radial basis functions to represent bond distances and spherical basis
functions to represent angles between bonds towards neighbouring
atoms.\cite{gasteiger2019directional} \\

\clearpage
\noindent
Many NN potentials are often additionally designed for application on
periodic systems including solids and crystals,\cite{schutt2018schnet}
or were updated to support periodicity.\cite{kaestner:2021} Others are
specifically designed to train on reference data to predict formation
energy, lattice parameters of the unit cells and other material
properties directly from the structural
fingerprint.\cite{zhang:2018,grossman:2018} The application of ML
(including NNs) to materials has been discussed in detail in recent
reviews\cite{ramprasad2017machine,wang2020machine,sutton2020identifying}
and is not further considered in the present work.  \\

\noindent
The field of NNs in computational chemistry has been and will continue
to be steadily developed to improve the capability and accuracy in
predicting reference data. In consequence, the selection of a model
should be done based on the problem at hand, the availability of the
code, its user friendliness, and the computational resources
available. It might not be necessary to use the most sophisticated
model if the task does not require that level of description.  Most of
the previously described architectures are based on open source NN
frameworks like Tensorflow\cite{tensorflow2015-whitepaper} or
PyTorch\cite{paszke2019pytorch} which open the possibility to
modifications and enhancements of the described models.\\

\section{Construction of PESs}
\label{sec4}
The collection of reference structures is an essential step in
constructing a molecular PES, especially since the underlying
functional form of the potential is not based on physical laws and is
inferred purely from reference data.\cite{behler2021four} Besides the
unfavourable scaling of the configurational space with system size,
the computational expense associated with a reference point is usually
high and depends on the level of quantum chemical theory used. Thus,
the number of expensive and non-trivial \textit{ab initio}
calculations needs to be restricted to a minimum and optimally covers
the configurational space most important/representative (this is an
open question in itself) to the problem at
hand.\cite{unke2021machine,bogojeski2020quantum} Ultimately, the
configurational space that is covered by the reference data set
defines the boundaries of application of the NNP. Therefore, knowing
the application(s) for which the PES will be used is essential when
generating the data.\\

\begin{figure}[b!]
\centering
\includegraphics[width=0.9\textwidth]{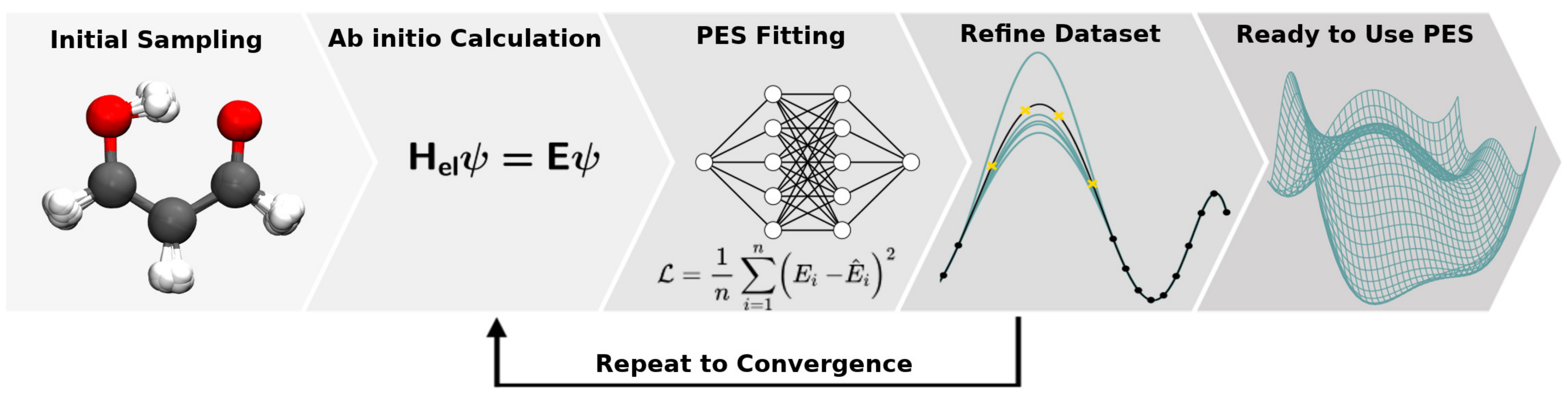}
\caption{The process of PES generation: The configurational space of a
  chemical system (here malonaldehyde) is sampled to obtain an initial
  set of geometries. A quantum chemical \textit{ab initio}
  calculation is carried out for each geometry to obtain reference
  data (including energies). After a NNP is fitted to the initial
  reference data set the resulting PES is validated thoroughly to find
  holes. New \textit{ab initio} calculations are run for scarcely
  sampled regions and a new NNP is fitted. These steps are repeated
  until the PES has the required quality before the PES can be used to
  study the chemical system.}
\label{fig:pes_generation}
\end{figure}

\noindent
Reference data sets can be generated using a multitude of strategies
which often requires the generation of an initial data set and
refining it iteratively. This iterative process is illustrated in
Figure~\ref{fig:pes_generation}.  Commonly employed strategies for
structure sampling, which are often combined, will be described in the
following. In addition to methods reviewed here, other possibilities
include Virtual Reality
sampling\cite{o2018sampling,amabilino2019training,amabilino2020training,chu2022exploring},
Boltzmann machines\cite{viguera2021} or sampling based on the AMONS
approach.\cite{lilienfeld2020slatm}

\subsection{Initial Sampling}
\subsubsection{Ab Initio MD}
\textit{Ab initio} MD (AIMD) constitutes an established means for
generating reference data that samples a part of the configuration
space of a chemical system.\cite{behler2021four} The temperature $T$
(or the velocities that are drawn from a Maxwell-Boltzmann
distribution corresponding to $T$) at which the simulation is run
determines which part of a PES is sampled, how strongly the molecular
geometries are distorted and whether or not reaction barriers are
crossed. If the chemical system under investigation has multiple
isomers, AIMD simulations can be run for all of them (partly) avoiding
the need of running a long simulation that samples all
isomers. Ideally, the sampling temperature $T$ is chosen to be higher
than the temperature at which the NNP is used. In other words, if the
reference data set that was used to train a NNP was generated at $T =
300$~K the NNP should not be used to run simulations at $T > 300$~K
because (most likely) configurations outside of the reference data set
are visited leading to a breakdown of the NNP. Thus, running AIMD at a
sufficiently high sampling temperature is needed to guarantee that the
production runs do not enter the extrapolation regime, while the lower
energy configurations are still sampled.\cite{unke2021machine}\\

\noindent 
The obvious disadvantage of running AIMD at the (final) level of
theory at which the reference data set is generated is the high
computational cost. This either limits the level of quantum chemical
rigor or it limits the extent to which the configurational space can
be sampled.\cite{gastegger2020molecular} Alternatively, configurations
can be generated using sampling by proxy.\cite{unke2021machine} This
approach involves running AIMD at a lower level of theory to sample
the PES and then perform single point \textit{ab initio} calculations
for a representative set of geometries at a higher level. This ideally
requires that the topologies of the lower and the higher level of
theory are similar to guarantee that the "correct" configurations are
sampled.  If the two PESs differ too much it is possible
that the regions explored on the lower level PES do not correspond to
relevant regions on the high level PES (which might happen if a force
field is used to guide the
sampling).\cite{behler2021four,unke2021machine} As a consequence, the
NNP could reach an extrapolation regime and exhibit a nonphysical
behaviour.\\

\noindent
Reactive chemical systems are usually associated with rare
events. When NNPs are used to study reactive systems it is, thus, not
sufficient to sample the reactant and product states since the
reaction path (which is rarely visited in a simulation) needs to be
part of the reference data set as well. TS regions can be sampled
using AIMD by employing a scheme similar to umbrella
sampling\cite{torrie1977nonphysical}, in which geometries around the
TS are sampled by harmonically biasing the molecule towards the TS.\\

\noindent
A simulation technique that is related to MD simulations and can be
used to generate configurations for the construction or refinement of
a reference data set is metadynamics.\cite{barducci2011metadynamics}
Converse to ordinary MD, metadynamics uses \textit{history dependent}
biasing potentials to artificially increase the potential of visited
regions on the PES and enhance the sampling of higher energy
regions.\\

\subsubsection{Normal Mode Sampling}
Normal mode sampling (NMS) was proposed to enable accelerated yet
chemically/physically relevant sampling of a PES.\cite{smith2017ani}
As the name suggests, NMS uses the normal modes of vibration of a
molecule to generate molecular geometries that cover configurational
space at which single point calculations can be carried out at a
desired level of theory. NMS is carried out as
follows\cite{smith2017ani}: i) the molecule of interest is optimized
at a desired level of theory ii) normal mode coordinates $Q = \{q_i\}$
(i.e. eigenvectors of the mass-weighted Hessian) and corresponding
force constants $K = \{k_i\}$ are determined (with $i \in [1,
  N_f=3N-5]$ or $i \in [1,N_f=3N-6$, for linear and non-linear
  molecules, respectively) iii) $N_f$ uniformly distributed random
  numbers $c_i$ with $\sum_i c_i \in [0,1]$ are generated iv)
  displacements for each normal mode are determined as $R_i = \pm
  \sqrt{\frac{3c_iN_Ak_b T}{K_i}}$ with $N_A$ and $k_b$ being the
  Avogadro number and the Boltzmann constant, respectively. This
  displacement is obtained by scaling an energy with $c_i$ [$E_k =
    \frac{3}{2}c_iN_Ak_bT$] and setting it equal to a harmonic
  potential [$U = \frac{1}{2} k r^2$]. v) determine the sign of the
  displacement $R_i$ randomly using a Bernoulli distribution to sample
  the attractive and repulsive parts of the potential vi) the
  normalized normal mode coordinates $q_i$ are scaled using $R_i$
  giving a new set of coordinates.\\

\noindent
Unlike the consecutive snapshots of an AIMD, NMS yields uncorrelated
molecular configurations in a very efficient manner. Nonetheless, the
sampling is based on a harmonic approximation of the potential well
and usually only geometries close to the respective equilibrium
structures are obtained. For larger displacements and large amplitude
motions, the harmonic approximation breaks down. Thus, NMS is often
used in conjunction with alternative sampling strategies or followed
by adaptive sampling.\cite{unke2021machine}

\subsubsection{Diffusion Monte Carlo}
Diffusion Monte Carlo (DMC) can be used to determine the zero-point
energy (ZPE) and wavefunction of a molecule by appropriately, yet
randomly, sampling the configurational
space.\cite{kosztin1996introduction}  The foundation of DMC is the
similarity of the imaginary time SE
\begin{align}
    \hbar\frac{\partial \Psi (x, \tau)}{\partial \tau} = \frac{\hbar^2}{2m}\nabla^2\Psi(x,\tau) - [V(x) - E_0]\Psi(x, \tau)
\end{align}
with the diffusion equation with a sink term allowing random-walk
simulations to estimate the ZPE and
wavefunction.\cite{li2021diffusion} Given a molecule, a set of walkers
is initialized (usually at some energy minimum), propagated randomly
at each time step $\tau$ and used to represent the nuclear
wavefunction. In one dimension, the displacement assigned to each of
the walkers is given by\cite{li2021diffusion}
\begin{align}\label{eq:dmc_disp}
    x_{\tau + \Delta\tau} = x_\tau + \sqrt{\frac{\hbar \Delta\tau}{m}r}
\end{align}
where $x_\tau$ corresponds to coordinates at time step $\tau$,
$\Delta\tau$ is the time step of the random-walk simulation, $m$
corresponds to an atomic mass and r is a random number drawn from a
Gaussian distribution, $\mathcal{N}(0,1)$. Once the walkers are
randomly displaced following Equation~\ref{eq:dmc_disp}, their
potential energy $E_i$ is determined. Based on $E_i$ with respect to a
reference energy $E_r$, a walker might stay alive, give birth to a new
walker or can be killed following the probabilities below:
\begin{align}
    P_{\rm death} = 1 - e^{-(E_i -E_r)\Delta\tau} \quad (E_i > E_r)\\
    P_{\rm birth} = e^{-(E_i -E_r)\Delta\tau} - 1 \quad (E_i < E_r)
\end{align}
Once the probabilities have been determined, the dead walkers have
been eliminated and new walkers are initialized, $E_r$ is adjusted
following
\begin{align}\label{eq:update_dmc}
    E_r(\tau) = \left<V(\tau)\right> - \alpha \frac{N(\tau) -
      N(0)}{N(0)}.
\end{align}
The averaged potential energy of the alive walkers is given by $
\left<V(\tau)\right>$, $\alpha$ governs the fluctuation in the number
of walkers and is a parameter, and $N(\tau)$ and $N(0)$ are the number
of alive walkers at time step $\tau$ and 0, respectively. The ZPE is
then approximated as the average of $E_r$ over all imaginary
time.\cite{kosztin1996introduction,li2021diffusion}\\

\noindent
The geometries sampled using the DMC scheme are physically meaningful
(the ensemble of walkers represents the nuclear ground state
wavefunction) and efficiently obtained by only using energies. In
comparison to AIMD, the DMC scheme has the advantage that it samples
configurations up to the ZPE, which becomes larger for bigger
molecules. The (quantum) exploration of a PES using DMC is typically
done after a first PES has been fitted and is used to refine the
reference data set.\cite{conte2020full} DMC has been proposed as a
tool to detect \textit{holes} (regions on a PES that have large
negative energies with respect to the global minimum) in ML based
PESs.\cite{conte2020full} These holes are caused by insufficient data
in specific regions in configuration space, for which a NNP without
any underlying physical knowledge leads to artifacts. As an
adaptation, DMC with artificially reduced masses has been proposed to
locate holes more efficiently due to the larger random displacements
(which are proportional to $1/\sqrt{m}$, see
Equation~\ref{eq:dmc_disp}).\\

\subsection{Validation and Refinement of the Data Set}
These holes were found to exhibit energies with large negative
values.\cite{nandi2019using} After an initial PES is fitted, a
thorough evaluation of the PES to discover any holes is needed. For
this reason, the family of active learning schemes which comprise
algorithms to systematically generate reference data sets have gained
considerable attention.\cite{shapeev2020active} The necessity for more
elaborate sampling schemes is related to the impracticality of an
exhaustive sampling of a PES and the high computational cost of
extensive \textit{ab initio} calculations. Typically, a first PES is
trained on reference data based on representative configurations. This
is followed by suitably extending the data set in an iterative fashion
in which similar configurations are avoided and configurations from
underrepresented regions of the PES are found and included into the
data set.\cite{shapeev2020active} This approach is usually termed
adaptive sampling (or on-the-fly
ML).\cite{csanyi2004learn,gastegger2017machine} Therefore, a
requirement for ML models to autonomously select new reference data is
the availability of an uncertainty estimation. If a defined
uncertainty threshold is exceeded for a particular configuration
electronic structure calculations are performed and used to extend the
reference data.\\

\subsubsection{Uncertainty Estimation}
\label{sec421}
Given the breadth of NN methods (or ML methods in general), various
approaches for uncertainty estimation exist. One of the most popular
methods is \textit{query-by-committee}.\cite{shapeev2020active} This
approach involves training/fitting a number of individual NNPs
(e.g. starting from different parameter initialization or on different
splits of the reference data set) and using the \textit{ensemble} for
predictions. In regions of the configuration space where sufficient
data is available the predictions of the different models agree
well. Conversely, the predictions for configurations for scarcely
sampled regions will diverge rapidly, and can be used to autonomously
select new configurations. A possible uncertainty metric for NNPs
is\cite{gastegger2020molecular}
\begin{align}
    \sigma_E = \sqrt{\frac{1}{\mathcal{N} - 1}
      \sum_{i=1}^\mathcal{N}(E_i - \overline{E})^2}
\end{align}
with $\mathcal{N}$ being the number of individual models, $E_i$ an
individual energy prediction and the average of all energy
predictions, $\overline{E}$. Similar metrics can certainly also be
adapted to other properties including the forces acting on the atoms
$\alpha$\cite{gastegger2020molecular}:
\begin{align}
    \sigma_F^\alpha = \sqrt{\frac{1}{\mathcal{N} - 1}
      \sum_{i=1}^\mathcal{N}\left\Vert F_i^\alpha -
      \overline{F}^\alpha\right\Vert^2}
\end{align}

\noindent
The use of \textit{query-by-committee} requires the training of
several independent models which incurs a high computational cost to
obtain the uncertainty. In addition to this, it has been found that
the uncertainty estimated by NNP ensembles are often
overconfident.\cite{Kahle2022} As a solution to this bottleneck,
methods that obtain the uncertainty in a single evaluation have been
proposed. Some us\cite{vazquez2022uncertainty} recently introduced a
modification of the PhysNet architecture that allows the calculation
of the uncertainty on the prediction through a method called
\textit{deep evidential regression}.\cite{amini2020deep} Using this
method, the energy distribution of the system is represented with a
Gaussian and its uncertainty as a gamma distribution. With this
approach, it is possible to obtain the prediction and the uncertainty
of the prediction in one single calculation. Other possibilities for
the prediction of uncertainties include the use of Bayesian NNs,
however, they imply a larger computational cost than the previously
described methods.\\

\subsubsection{Elaborate Sampling Techniques}
With the availability of an uncertainty measure and an initial PES,
geometries from underrepresented regions on the PESs can easily be
identified: The initial PES is used to guide the sampling of new
structures ((by MD, DMC, metadynamics, ...) and if the uncertainty
measure (e.g. $ \sigma_E$) exceeds a threshold, \textit{ab initio}
calculations are performed for the geometry and the data set is
suitably extended. These more systematic approaches of generating
reference data sets offer a number of advantages over random
methods. Since including similar configurations is avoided and new
data is only added for scarcely sampled regions, the approaches are
clearly more data efficient requiring less expensive quantum chemical
computations. Additionally, since the NNP that is used to guide the
sampling of new geometries is topologically very similar to the
\textit{ab initio} PES it is assured that configurations, that are
similar to the configurations visited in AIMDs, are sampled. The
quality of the uncertainty estimate is crucial for all adaptive
sampling schemes. While an over-confident estimate leads to an
inaccurate PES (in the worst case holes are overlooked) an
under-confident estimate leads to the inclusion of redundant
configuration and unnecessary, computationally expensive \textit{ab
  initio} calculations. Zipoli and coworkers report that adding new
configurations based on uncertainty estimation from an ensemble of
NNPs does not show significant differences from random
sampling.\cite{Kahle2022} Contrary to that,
Pernot\cite{pernot2022prediction} and Zheng \textit{et
  al}\cite{zheng2022toward} find that querying the uncertainties from
ensembles are well suited for outlier detection and adaptive sampling.
This clearly indicates the necessity for future studies exploring more
elaborate sampling techniques.

\section{Knowledge Transfer}
\label{sec5}
Most ML algorithms (foremost deep learning) heavily rely on abundant
training data to extract the underlying patterns in very complex
data. This severe \textit{data dependence} is one of the major
drawbacks to deep learning.\cite{tan2018survey} The collection of big
data sets is a cumbersome and expensive task impeding the generation
of large, high-quality data sets. While this time-consuming endeavor
might be possible for some areas of application (e.g. manually
labeling images for an image recognition task) insufficient training
data/data scarcity is an inevitable problem in other domains
(e.g. drug discovery).\cite{tan2018survey,cai2020transfer} Thus,
transfer learning (TL)\cite{pan2009survey,tan2018survey} and related
approaches including $\Delta$-ML\cite{fu:2008,DeltaPaper2015},
dual-level Shepard interpolation,\cite{nguyen1995dual}
multifidelity learning\cite{batra2019multifidelity} or the multilevel
grid combination technique\cite{zaspel2018boosting} have been proposed
to circumvent the severe \textit{data dependence/scarcity} or
expensive labeling efforts by knowledge transfer. Thereby, exploiting
the knowledge acquired by solving one task (a \textit{source} task) to
solve a new, related task (a \textit{target} task) forms its common
ground.\cite{pan2009survey}\\

\noindent
Besides addressing the data scarcity dilemma, knowledge transfer also
helps reducing training times, computer resources (which both are
significant for large data sets/models\cite{hinton2015distilling}) and
their energy consumption. Recently, the CO$_2$ emission for training
common natural language processing (NLP) models has been studied,
which, depending on their size, can exceed a car's lifetime CO$_2$
emission.\cite{strubell2019energy}\\

\begin{figure}[h!]
\centering
\includegraphics[width=0.9\textwidth]{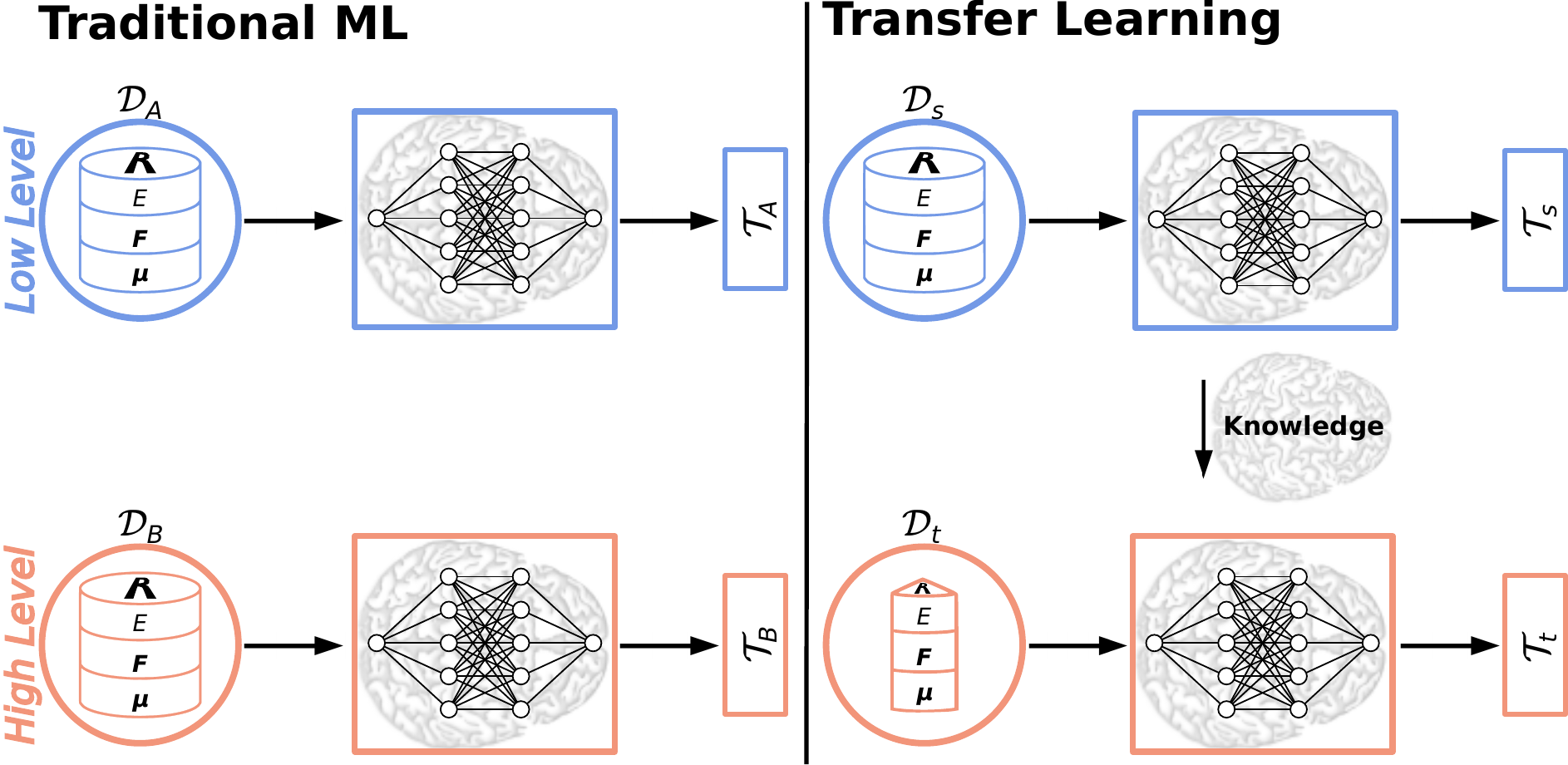}
\caption{Illustration of the difference between traditional ML and TL
  approaches. In traditional ML, two different models are trained for
  two different tasks ($\mathcal{T}_A$ and $\mathcal{T}_B$), although
  the two tasks might be related (e.g. predicting the MP2 and the
  CCSD(T) energy of a given configuration). In TL, however, the
  knowledge gained from solving a source task ($\mathcal{T}_s$) in the
  source domain ($\mathcal{D}_s$) is used to solve a target task
  ($\mathcal{T}_t$) (e.g. by fine-tuning the weights and biases). In
  the context of PES generation, typically a (global) PES is developed
  at a low level of theory and then transfer leaned with less data
  calculated at a considerable higher level of theory (e.g. CCSD(T)).}
\label{fig:tl_domains}
\end{figure}

\noindent
Traditional ML problems usually proceed in a domain $\mathcal{D}$ and
try to solve a specific task $\mathcal{T}$. In the context of
molecular PESs, the domain $\mathcal{D}$ is a set of molecular
configurations (defined by $\{\mathbf{R}, \mathbf{Z}\}$) with their
associated descriptors (see Section~\ref{sec:descriptors}) and the
task involves the prediction of the corresponding energies
$E^{BO}_\lambda(\mathbf{R})$ (Equation~\ref{eq:bo-pes}).  Considering
two domains (a source $\mathcal{D}_s$ and a target domain
$\mathcal{D}_t$) and two learning tasks ($\mathcal{T}_s$ and
$\mathcal{T}_t$) from the perspective of traditional ML, two separate
machines are trained to solve the two tasks (see
Figure~\ref{fig:tl_domains}). In contrast, TL circumvents learning to
solve both tasks from scratch by facilitating the learning of
$\mathcal{T}_t$ with knowledge from $\mathcal{T}_s$ (see
Figure~\ref{fig:tl_domains}).  Here, the domains and/or tasks can
differ for TL giving rise to three distinct
cases.\cite{pan2009survey,cai2020transfer} i) The domains are the
same, $\mathcal{D}_s = \mathcal{D}_t$, while the tasks differ,
$\mathcal{T}_s \neq \mathcal{T}_t$. This situation can, e.g., be found
for TL between molecular properties (\textit{inductive learning}) ii)
The domains differ, $\mathcal{D}_s \neq \mathcal{D}_t$, while the
tasks remain the same $\mathcal{T}_s = \mathcal{T}_t$. This
corresponds to \textit{transductive learning} and can be found for TL
between different molecular data sets. iii) Both, the domains and the
tasks differ, $\mathcal{D}_s \neq \mathcal{D}_t$ and $\mathcal{T}_s
\neq \mathcal{T}_t$. All three subsettings have in common that they
try to learn/improve the target predictive function $f_t(\cdot)$ of
$\mathcal{T}_t$ in $\mathcal{D}_t$ using the knowledge in
$\mathcal{D}_s$ and $\mathcal{T}_s$ which is the definition of
TL.\cite{pan2009survey}\\

\noindent
The training of NNPs typically requires thousands to tens of thousands
of \textit{ab initio} calculations even for moderately sized
molecules, which often limits the quantum chemical calculations to the
level of density function theory (DFT).  If highly accurate molecular
properties are needed, researchers usually resort to the coupled
cluster with perturbative triples (CCSD(T)) level of theory. This
``gold standard'' - CCSD(T) - scales as $N^7$ (with $N$ being the
number of basis functions)\cite{friesner2005ab}, which makes
calculating energies and forces for large data sets and larger
molecules impractical. Thus,
TL\cite{smith2017ani,mo2020transfer,kaser2020reactive,kaser2021vpt2,kaser2022transfer}
and related $\Delta$-learning
approaches\cite{DeltaPaper2015,nandi2021delta,qu2021breaking,qu2022delta}
gained a lot of attention in recent years and were shown to be data
and cost effective alternatives to the "brute force" approach in
quantum chemistry: A low level PES based on a large data set of cheap
reference data (e.g. DFT) is generated first, which then is used to
obtain a high level PES based on few, well chosen high level of theory
(e.g. CCSD(T)) data points.\\

\subsection{Deep Transfer Learning}
Deep TL\cite{cai2020transfer} combines deep NN architectures with TL
among which fine-tuning is the most commonly used
technique. Fine-tuning, which is a parameter-based TL technique,
assumes that the weights and biases of a deep NN that was trained on a
source task $\mathcal{T}_s$ contain useful information to solve a
(related) target task $\mathcal{T}_t$. In the context of molecular
PESs, a lower level (LL) PES is obtained by training a deep NN on a
large data set of energies/gradients determined at a low level of
theory. Then, the parameters (weights and biases) of the LL
PES are migrated to the target model for which they serve as the
initialization (a good initial guess). The target model (i.e. the
transfer learned model) is then fine-tuned (retrained) on a small data
set of high-level of theory energies/gradients. The fine-tuning
technique that migrates the parameters of a LL PES to a high level
(HL) PES is shown in Figure~\ref{fig:tl_domains}.\\

\noindent
There are certain subtleties when applying TL in practice. TL can be
performed without any further restriction to the fine-tuning for which
all weights and biases are allowed to adapt to the new HL
data. Conversely, it is possible to fix the weights and biases of
particular layers. Usually, the first hidden layers are fixed and only
the last layer(s) are allowed to adjust (alternatively a new, final
layer can be added keeping the LL model as is). Fixing a portion of
the NN parameters limits its flexibility but might help in reducing
overfitting for small data sets. Recently, TL in combination with NNs
was used for structure-based virtual screenings of
proteins.\cite{imrie2018protein} The authors found that fine-tuning a
full NN worked best for kinases, proteases and nuclear proteins,
however, fine-tuning only the final layer yielded better results for
G-protein-coupled receptors (GPCRs). They speculate that this is
caused by the limited and less diverse data for GPCR targets. Besides
the need to avoid overfitting, it is imaginable that for NNs that
employ learnable descriptors of the atomic/molecular configuration it
might be beneficial to freeze the parameters that are used to learn
the descriptor for the fine-tuning step. Instead of freezing a portion
of the layers, fine-tuning with differential learning
rates\cite{mishra2019improving} (i.e. having different learning rates
for different parts of the NN) could allow minimal changes to early
layers (e.g. where the descriptors are learned) and larger adjustments
to the later layers. Although empirical rules are followed in the
community, accepted criteria for choosing TL methods are essentially
nonexistent.\cite{cai2020transfer} \\

\subsection{$\Delta$-Machine Learning}
The $\Delta$-machine learning approach was developed in the context of
kernel-based methods and is motivated by the fact that the heaviest
burden in quantum chemical calculations is the determination of a tiny
energy contribution to a (approximate) total
energy.\cite{DeltaPaper2015} The approximate energy often is able to
describe the general chemistry/physics of a given system, while the
determination of the "$\Delta$" comes at a tremendous computational
cost due to adverse scaling with system size of correlated electronic
structure methods. For a molecular property, the $\Delta$-ML
prediction is modeled as a LL value plus a correction towards a HL
value following
\begin{align}\label{eq:deltaml_2015}
P_{\rm HL}(R_{\rm HL}) \approx \Delta_{\rm LL}^{\rm HL}(R_{\rm LL}) =
P'_{\rm LL}(R_{\rm LL}) + \sum_{i=1}^N \alpha_ik(R_{\rm LL}, R_{\rm
  i}).
\end{align}
The high level property $P_{\rm HL}$ (e.g. enthalpy $H_{\rm HL}$) at a
relaxed molecular geometry $(R_{\rm HL})$ is approximated as a related
property $P'_{\rm LL}(R_{\rm LL})$ (e.g. energy $E_{\rm LL}$) obtained
at the LL plus a correction term\cite{DeltaPaper2015} that is obtained
from ML (Reference~\citenum{DeltaPaper2015} employed Slater type basis
functions $k$ and kernel ridge regression (KRR) to obtain the
regression coefficients $\alpha_i$). The $\Delta$-ML approach as
defined in Equation~(\ref{eq:deltaml_2015}) allows modeling changes in
level of theory (e.g. DFT $\rightarrow$ CCSD(T)), molecular property
(e.g. energy $\rightarrow$ enthalpy) and molecular geometry.  Although
the $\Delta$-ML approach is often used in conjunction with
kernel-based methods, a \textit{correction PES $\Delta$} (i.e. $V_{\rm
  HL} = V_{\rm LL} + \Delta$) can also be learned using
NNs\cite{liu2022permutation}. The resulting HL PES $V_{\rm HL}$ can
either be used directly (requiring the evaluation of two models) or
can be used as a proxy to generate a larger data set for a final
training containing many, though approximate, HL
points.\cite{liu2022permutation} As is common for the ML field,
different flavours of $\Delta$-ML
exist.\cite{DeltaPaper2015,zaspel2018boosting,batra2019multifidelity,zhu2019artificial,dral2020hierarchical,bogojeski2020quantum,nandi2021delta,qu2021breaking,ruth2022machine,liu2022permutation}\\

\noindent
Recent work proposed "$\Delta$-DFT" 
that uses Kohn-Sham (KS) electron densities $\rho^{\rm KS}$
to correct the DFT energy towards, e.g., a coupled cluster energy 
following
\begin{align}
    E^{\rm CC} = E^{\rm DFT}[\rho^{\rm KS}] + \Delta E[\rho^{\rm KS}]
\end{align}
using KRR.\cite{bogojeski2020quantum} While the formalism of DFT and
wavefunction based approaches (such as CCSD(T)) differ radically (also
note that the CCSD(T) density is not routinely calculated and not
needed to obtain the CCSD(T) energy), the "learnability" of DFT and
CCSD(T) energies from KS densities was studied alongside the
$\Delta-$DFT approach. The authors find starting from $\rho^{\rm KS}$
learning DFT and CCSD(T) energies directly is associated with
approximately the same effort. However, learning 
$\Delta E[\rho^{\rm KS}]$ was more efficient and yielded lower 
out-of-sample errors at smaller training set sizes.\cite{bogojeski2020quantum}

\section{Exemplary Applications of NNPs in Molecular Simulations}
The high flexibility of NNs allows the representation of PESs for a
wide range of chemical systems and reactions as long as a sufficiently
large reference data set is available from \emph{ab initio}
computations at a sufficient level of theory to correctly describe the
physics in the system. This section presents several typical
applications of NNPs in molecular simulations.

\subsection{Gas Phase Spectroscopy}
In a recent review, Manzhos and Carrington report advances of NNPs
and applications in classical and quantum dynamics of small and
reactive systems.\cite{carrington2021rev} They point out that for
small systems modern NNPs are still outperformed by
permutationally invariant polynomial
(PIP\cite{bowman2009pip,bowman2022pip}) methods in terms of PES
fitting error which, however, does not translate to significant
deviations in computed observables such as vibrational
frequencies.\cite{majumder2015pipvsnn} As an example, the RMSE of a
Gaussian process regression (GPR) model potential (0.017\,kcal/mol,
5.98\,cm$^{-1}$) is half of that of a NNP (0.034\,kcal/mol,
12.03\,cm$^{-1}$) with regard to 120\,000 reference points for
formaldehyde.  However, the RMSE of the first 50 (100) predicted
vibrational frequency levels with respect to their reference is
0.43\,cm$^{-1}$ (0.82\,cm$^{-1}$) for the NN and 0.46\,cm$^{-1}$
(0.82\,cm$^{-1}$) for the GPR potential.  When the potential models
are fitted to a subset of reference points with high significance for
the vibrational frequency prediction, the RMSE of the the first 50
(100) predicted vibrational frequency levels differs substantially
with 0.21\,cm$^{-1}$ (0.30\,cm$^{-1}$) for the NN and only
0.04\,cm$^{-1}$ (0.06\,cm$^{-1}$) for the GPR
model.\cite{kamath2018nnvsgpr,carrington2021rev} \\

\begin{figure}[h!]
\centering
\includegraphics[width=0.9\textwidth]{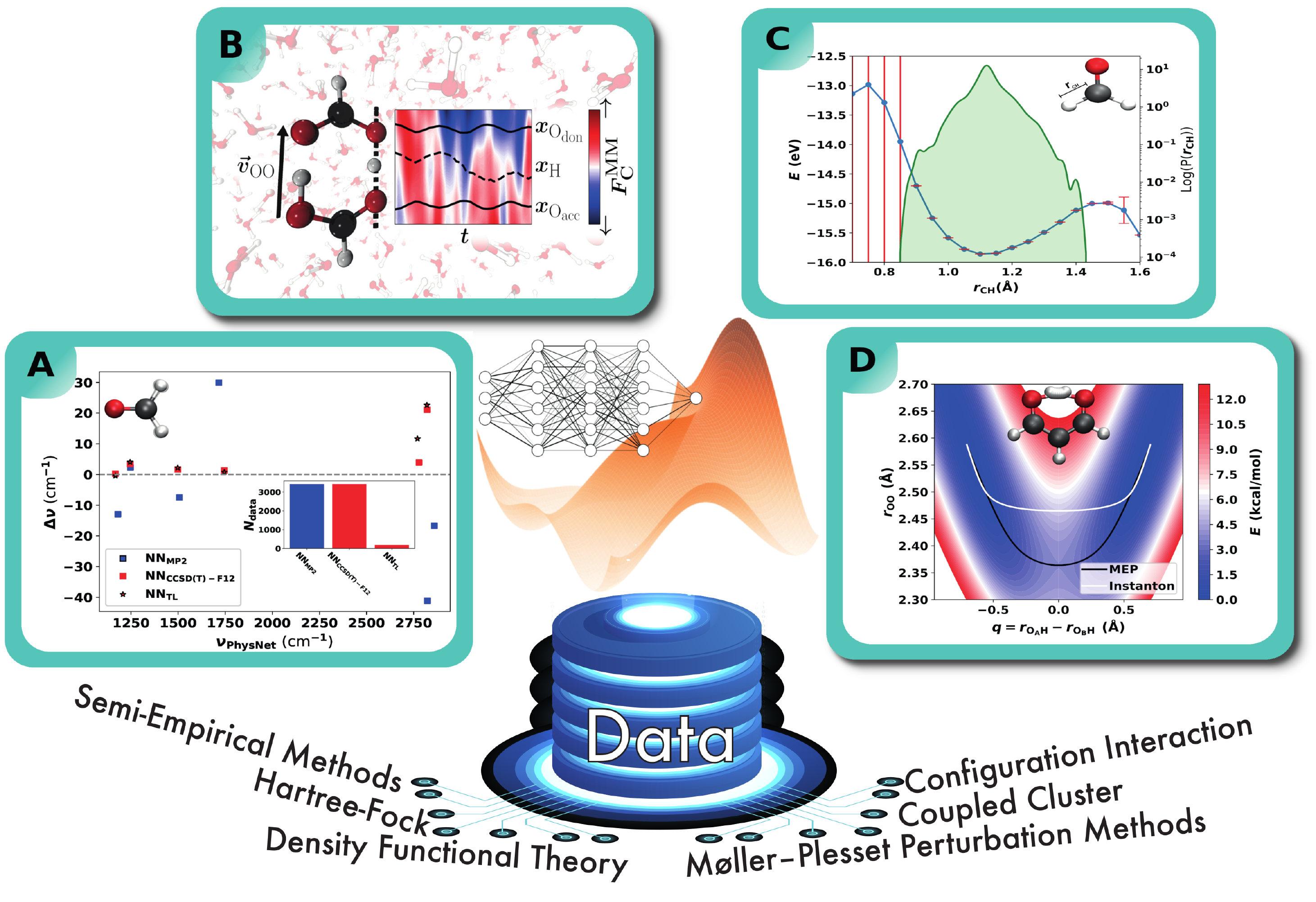}
\caption{Schematic representation of the exemplary applications of
  NNPs. A: Performance of a NNPs based on MP2/aVTZ and
  CCSD(T)-F12/aVTZ-F12 with respect to experiment. NNPs trained from
  scratch are compared to the more-data efficient TL approach and the
  anharmonic frequencies are obtained from VPT2
  calculations.\cite{kaser2021vpt2} B: Double proton transfer in
  formic acid dimer from mixed ML/MM/MD
  simulations.\cite{toepfer2022fad} The time series next to the
  molecular structure shows the variation in the background solvent
  field depending on time across one proton transfer event.  C: 1D cut
  of the PES of the C-H bond in formaldehyde (upper right) calculated
  with the PhysNet evidential model (blue curve). Red bars indicate
  the predicted variance by the model. The green distribution shows
  the logarithm of the probability distribution of the distances
  covered by the training set. D: The two-dimensional projection of a
  NN-trained PES of CCSD(T) quality for proton transfer in
  malonaldehyde. The white and black traces are the instanton and
  minimum energy paths, and the PES is used to calculate tunneling
  splittings.\cite{kaeser2022tlrpi}.}
\label{fig:compound}
\end{figure}

\noindent
The application of NNPs to determine anharmonic vibrational
frequencies in combination with TL has been studied in
Reference~\citenum{kaser2021vpt2}. For that purpose, a NN of the
PhysNet type is trained on \textit{ab initio} energies, forces and
dipole moments and employed in second order vibrational perturbation
theory (VPT2) calculations that are directly compared to their
experimental counterpart. A total of eight molecules are studied from
which the results for formaldehyde are shown in
Figure~\ref{fig:compound}A as it allows a good comparison of a TL
scheme with a model that is trained "from scratch" due to its small
size. A PhysNet model that is trained on MP2 data (NN$_{\rm MP2}$)
yields errors up to 40~cm$^{-1}$ with respect to the experimental
values, while the CCSD(T)-F12 model (NN$_{\rm CCSD(T)-F12}$) has a
maximum deviation of $\sim 20$~cm$^{-1}$. Both NN$_{\rm MP2}$ and
NN$_{\rm CCSD(T)-F12}$ were trained on roughly 3400 \textit{ab initio}
energies, forces and dipole moments, 
for which the computation at the CCSD(T)-F12 level of theory 
requires high computational effort.
In contrast, 6\% of the CCSD(T)-F12 reference points are sufficient 
to transfer learn a NN$_{\rm MP2}$ model and achieve an accuracy that is 
within $\sim 7$~cm$^{-1}$ of NN$_{\rm CCSD(T)-F12}$ trained on the full 
reference set from scratch.\\

\subsection{Condensed Phase Simulations}
Even though NNPs scale more favourably with the number of atoms, the
construction of a reference data set for molecular compounds still
requires several thousand \emph{ab initio} calculations. As NNPs are
mathematical representations of the input data and are uninformed
about the underlying physics governing intermolecular interactions,
their extrapolation capabilities are rather limited. This also
concerns the transferability of NNPs optimized on smaller molecular
clusters towards larger clusters or even periodic systems. This issue
has been addressed recently, for instance, by K\"astner and coworkers
on liquid water and Marx and coworkers on protonated water clusters
using NNPs.\cite{kaestner2022pbc,marx2021zundel}\\

\noindent
K\"astner and coworkers train a Gaussian moment NN (GM-NN) model on
DFT rev-PBE-D3 reference data of water cluster configurations produced
by \emph{ab initio} MD simulation at 150, 300 and 800\,K, and study
its transferability to a periodic bulk water system with 64 molecules
from \emph{ab initio} MD simulation at
400\,K.\cite{kaestner2022pbc,kaestner2020gmnn} The GM-NN model trained
on clusters containing 30 to 126 water molecules can reproduce the
total energy of the periodic bulk water system well, although with a
slightly broader error distribution as for the model trained on the
periodic system. The potential energy predicted by the cluster model
for the periodic systems are also arbitrarily shifted mainly due to
the differences in the non-periodic and periodic computational system
setup. MD simulation of a periodic water box at 300\,K with the model
potentials trained on clusters (cluster model) and periodic reference
data (bulk model) produce radial distribution function that agree well
and X-ray diffraction spectra are close to experimental ones. The
computed water molecule self-diffusion coefficients and equilibrium
density from simulations with the cluster model are about 18\,\%
larger ($2.15\cdot10^{-9}$\,m$^2$/s and $1.02$\,g/cm$^{3}$) than with
the bulk model ($1.82\cdot10^{-9}$\,m$^2$/s and $0.86$\,g/cm$^{3}$)
but closer to the respective experimental values
($2.41\cdot10^{-9}$\,m$^2$/s and $1.00$\,g/cm$^{3}$).  Detached from
the evaluation of the rev-PBE-D3 method and MD setup to accurately
reproduce experimental water properties, the case study shows
transferability of the cluster model to reproduce bulk properties.
However, the authors mention that further studies are necessary to get
insights into the deviation in the computed properties of both models
as both water cluster and periodic water system are based on the same
physical-mathematical description.  Only water molecules closer to the
cluster surface experience different strain energy than bulk water due
to the lack of bonding partners.  \\

\noindent 
Great transferability is also shown by Marx and coworkers using a
HDNNP model trained on protonated water cluster H$^+$(H$_2$O)$_n$
($n=1-4$) with up to four water molecules to representing the PES of a
protonated water hexamer
H$^+$(H$_2$O)$_6$.\cite{marx2021zundel,behler2017first} The reference
data for the protonated water clusters $n=1-4$ were produced by an
automatic fitting procedure that performs DFT based \emph{ab initio}
MD and path integral MD (PIMD) simulation at 1.67, 100 and 300\,K to
sample relevant configurations.  Within a repeated fitting procedure,
holes in the reference data set are detected by estimating the
uncertainty as described in section \ref{sec421} or configurations
were included where the local descriptors (ACSFs) of configurations in
the MD simulation leave the range of the reference data
set.\cite{marx2020sampling} A final data set is created from reference
data of the configurations computed at CCSD(T*)-F12a/aug-cc-pVTZ level
of theory. Extrapolation of the NN model trained on the smaller
cluster $n=1-4$ to configuration of the protonated water hexamer
yields a mean absolute energy error about three times higher than for
the original training data set that is 0.026, 0.031, 0.038\,kcal/mol
(0.11. 0.13, 0.16\,kJ/mol) per atom against 0.007, 0.010,
0.012\,kcal/mol (0.03. 0.04, 0.05\,kJ/mol) per atom from the sampling
procedure at 1.67, 100 and 300\,K, respectively.\cite{marx2021zundel}
Again, an arbitrary shift is added to the predicted energies of the
hexamer to minimize the error between the predicted and the reference
energies.  The ability to extrapolate is illustrated by comparing the
potential energy sequence for 25\,fs between an \emph{ab initio} MD
and the MD simulation using the NNP.  It is further noticeable, that
the extrapolation towards the hexamer potential failed in PIMD
simulations for which unphysical configurations are reached if the NNP
is trained only on tetramer configurations ($n=4$). The authors
conclude that the transferability towards larger cluster sizes
improves if smaller clusters are included within the training data
set.  \\

\subsection{Reaction Rates}
The reaction of methane with molecular oxygen is one of the most
fundamental but highly complex combustion processes involving more
than one hundred different reaction steps as shown by
experiments.\cite{grimech} Zhu and Zhang report MD results of the
combustion reaction including 100 methane and 200 oxygen molecules at
3000\,K simulated for 1\,ns.\cite{zeng2020complex} They used the
DeepMD model potential that was fitted to reproduce 578\,731 reference
DFT energies at the MN15 level of
theory.\cite{wang2018deepmd,zhu2014ch4o2ref} In their simulation they
detected 505 molecular species and 798 different reactions where 130
reaction steps are also reported from experiments.\cite{grimech} A
selection of computed reaction rates deviates from experiment by up to
two orders of magnitude, but combustion reactions usually involve the
formation of radical species, that might require a non-adiabatic
molecular dynamics approach which are highly non-trivial.  \\

\noindent
Marquetand and coworkers applied the SchNarc approach to investigate
the photodissociation reaction of tyrosine that shows a dissociation
channel of a hydrogen radical with a chemically non-intuitive path
which is called roaming.\cite{marquetand2022roaming} Roaming was
originally explored experimentally and computationally in formaldehyde 
by Bowman and coworkers in 2004 but real-time experimental observation
were not achieved until 2020.\cite{bowman2004roaming,endo2020roaming} 
The NNP is learned to
reproduce 29 energy values and force values for electronic singlet and
triplet states and 812 spin-orbit couplings. They simulated over 1000
trajectories of at least one picosecond which, in comparison, would
take over eight years for \emph{ab initio} MD simulation on a
high-performance computer.  About 17\,\% of the trajectories show the
roaming of the hydrogen atom in photoexcited tyrosine that lead to a
higher ratio of subsequent further fragmentation than in non-roaming
trajectories. This application marks a major step forward towards
atomistic simulations of photoexcitation reactions in larger molecules
like proteins that lead to further insight in, e.g., photosynthesis,
harmful photodegradation or drug designing for phototherapy.  \\

\subsection{Hybrid ML/MM Simulations of Solvated Systems}
The use of NNPs as force fields promotes the performance of MD
simulations in comparison to the \emph{ab initio} MD counterpart. But
even if the computational cost of NNPs scales by a similar factor of
$\sim O(N^{1-2})$ as empirical force fields do, due to their more
compact and explicit functional form empirical force fields are
considerably more efficient in general. Thus, a significant speed-up
in MD simulations can be achieved by decomposing the force field into
a contribution from a NNP (ML part) for, e.g., a solute of interests
or a reactive center in a protein, an empirical force field (MM part)
for solvent molecules or protein backbone structures, and a coupling
(or embedding) between the ML and MM parts. This approach is well
known and applied in QM/MM MD simulations.\cite{hu2008qmmmreview} \\

\noindent
One straightforward approach was pursued to investigate the double
proton transfer reaction in cyclic formic acid dimers and the
electrostatic impact of a water solvent on the reaction rate as shown
in Figure~\ref{fig:compound}B.\cite{toepfer2022fad} Here, a 
PhysNet model was trained with a
reference data set including formic acid dimers and monomers in the
gas phase at MP2/aug-cc-pVTZ level of theory. 
The model accurately reproduces the energies, forces and
molecular dipole by assigning atom centered
charges.\cite{unke2019physnet} The interaction potential between
formic acid and the TIP3P water solvent consists of Lennard-Jones
terms with parameters from the CGenFF\cite{mackerell2010CgenFF} 
force field and electrostatic interactions between the atom charges 
from the TIP3P \cite{jorgensen1983tip3p} 
water atoms and the configurational dependent PhysNet
charges of the formic acid atoms. The advantage is the lower
computational cost to produce trajectories with lengths of multiple
nanoseconds to statistically sample the raw double proton transfer
events with a rate of just 1\,ns$^{-1}$ at 350\.K.  Furthermore, the
NNP fit inherently includes the coupling of the reactive potential
path of the proton transfer with other structural dependencies such as
the C-O bond order of the acceptor and donor oxygen and the dimer
dissociation reaction into formic acid monomers. On the other hand,
such an approach does not include the mutual polarization of the
formic acid charges and the water solvent which, in the present
case, is however expected to be small. This is akin to a mechanical
embedding known from QM/MM schemes.\cite{ho2018embedding} \\

\noindent
Applications of electrostatic embedding in ML/MM simulation are
reported by Riniker and coworkers as well as Gastegger and
coworkers.\cite{riniker2021mlmm,gastegger2021mlsolv} Here, the ML-MM
interaction potential includes the polarization of the ML system by
the electric field originating from the MM compounds. Riniker and
coworkers modified the HDNNP by providing two sets of local
descriptors for just ML solute atoms and surrounding MM solvent atoms,
separately. The model is trained to reproduce either the ML atom
potential and the electrostatic component of the ML-MM atom
interaction itself (pure ML/MM) or in accordance of the
$\Delta$-learning approach an energy correction of both components to
improve from computational cheap tight-binding DFT result towards more
accurate reference data
((QM)ML/MM).\cite{riniker2021mlmm,behler2015constructing} This
approach demands larger reference data sets from QM calculations to
sample solute configurations with different solvent distribution where
the solvent is represented as their respective MM point charges.
However, the $\Delta$-learning (QM)ML/MM approach applied to
tight-binding DFT computations have been shown to achieve higher
accuracy even with fewer reference samples than the pure ML/MM
model.\\

\noindent
The accuracy is illustrated by running \textit{NPT} simulations of
S-adenosylmethionate and retinoic acid in explicit water solvent at
298\,K and 1\,bar using the pure ML/MM and the (QM)ML/MM model for
5000 and 2000 integration steps of 0.5\,fs, respectively, and
comparing it to reference QM/MM results.\cite{riniker2021mlmm} The
mean absolute error for the (QM)ML/MM model is up to one magnitude
lower with 1.4\,kcal/mol (5.8\,kJ/mol) and 12.6\,kcal/mol
(52.8\,kJ/mol) than the pure ML/MM model with 4.3\,kcal/mol
(18.1\,kJ/mol) and 17.9\,kcal/mol (74.9\,kJ/mol). One integration step
with the (QM)ML/MM model takes less than a second on 1 CPU while the
reference QM/MM model at DFT BP86/def2-TZVP level is about 3
magnitudes slower with about 60 to 80 minutes on 4 CPUs. A potential
disadvantage of the (QM)ML/MM model is that certain solute
configurations at the tight-binding DFT level may fail to converge or
converge only slowly, e.g., during a reaction.  \\

\noindent
Gastegger and coworkers presented the FieldSchNet model, a
modification of the SchNet model that includes energy contributions
from interactions between predicted atomic charges and dipoles, but
also with an external field such as the electric field originating
from a set of point
charges.\cite{gastegger2021mlsolv,schuett2017schnet} The advantage of
such elaborated models is the sensitivity of the potential energy to
changes in atomic positions, electric and magnetic fields that enable
the computation of response properties such as forces, molecular
dipole moments, polarizabilities, and atomic shielding tensors that
are crucial for the direct prediction of, e.g., IR, Raman and NMR
spectra. As the atomic charges and dipoles of the ML treated system
respond to the external field caused by MM atoms point charges, this
model is considered to be electrostatic embedding.  Consequently, it
has the same requirement for additional sampling of ML system
configurations in different arrangements of MM atomic point charges as
the model of Riniker and coworkers described above.  \\

\noindent
For ethanol in vacuum, PIMD simulations with FieldSchNet yield
excellent agreement in terms of frequency shifts and widths between
predicted IR/Raman spectra and experimentally measured ones. For
liquid ethanol, IR spectra were predicted from MD trajectories with an
explicit ML/MM solvent model of one ML treated ethanol molecule in a
MM treated ethanol solvent.  The explicit ML/MM approach shows great
agreement with experimental IR spectra in the low frequency region and
a blue shift for the C-H and O-H stretch vibrations bands in the high
frequency range due to missing anharmonicity effects by the MD
approach.  MD simulations with an implicit PCM solvent model do not
yield an IR spectra with significant differences from gas phase
spectra as it fails to capture hydrogen bridging between ethanol
molecules.\cite{mennucci2012pcm} However, the applied ML/MM model
still predicts the intermolecular ML-MM potential between ML ethanol
and the MM solvent by the CGenFF\cite{mackerell2010CgenFF} force field
with fixed atomic charges.  The implementation of the electrostatic
interaction between predicted atomic charges and dipoles by
FieldSchNet and the MM point charges is a highly non-trivial task and
would further increase the computational costs. It limits the
application range to systems where the the ML-MM interaction potential
is sufficiently well described by the MM force field that may not work
for dynamics with complex configurational changes or chemical
reactions. \\

\noindent
Electrostatic embedding in the QM/MM approach (and the ML/MM
approach)\cite{riniker2021mlmm} includes the QM-MM electrostatic
interaction and the polarization of the QM system by the electric
field of the MM atoms but not vice versa.  The highly expensive task
to approximate the polarization of the MM system by the electric field
of the QM system is part of polarizable embedding
schemes.\cite{mennucci2020embedding} An analogue for the hybrid ML/MM
model is developed Westermayr, Oostenbrink and coworkers with their
buffer region NN approach (BuRNN).\cite{oostenbrink2022burnn} Here, a
buffer region around the ML atoms is defined by a cutoff sphere to
select MM atoms within the sphere.  The ML and selected MM atoms are
the input to a modified SchNet model to predict the potential energy
between the ML atoms, the ML-MM interaction energy and a polarization
correction energy to the classical MM potential of the MM atoms within
the buffer sphere to match reference potential data.  The modified
SchNet model also predicts atomic point charges for the ML atoms and
MM atoms within the buffer region, which are used to compute the
electrostatic interaction to the remaining MM atoms in the system
outside the buffer region.  The potential energy of the atoms in the
inner region are predicted by a modified SchNet model.  As for
electrostatic embedding, potential energy and charge distribution of
the ML system are impacted by the MM atoms within a buffer region and,
additionally, interaction energy and atomic charges of the respective
MM atoms are impacted by the ML system.  A major disadvantage is the
high computation cost for the reference data set, that requires two
quantum electronic calculation for configuration samples of (1) the ML
system and MM atoms in the buffer region and (2) the MM atoms in the
buffer region alone to predict the polarization correction term.  \\

\noindent
The BuRNN approach was applied to a hexa-aqua iron(III) complex
simulated by a ML treated Fe$^{3+}$ ion in a water solvent described
by the SPC model.  A buffer region was defined by a cutoff radius of
5\,\AA\ around the Fe$^{3+}$.  MD simulation of 10\,ns shows smooth
diffusion of water molecules entering and leaving the buffer region
and reveal power spectra that match the low frequency bands around
180, 310 and 500\,cm$^{-1}$ observed in experiments very well. Radial
and improper and distributions between Fe$^{3+}$ and the oxygens of
the coordinated water match with distributions from QM/MM simulation
with electrostatic embedding and are within experimental
estimations.\\

\noindent
All the presented applications show an active field of developments in
hybrid ML/MM approaches towards accurate MD simulation of solutes or
reactive species in the presence of a solvent. A major gain in
computational efficiency and much longer simulation times at
comparable accuracy are achieved by replacing QM methods with a NNP.
However, the effort to generate a reference data set that sufficiently
samples the relevant configurational space of the ML system in
combination with different solvent configuration depends significantly
on the embedding scheme. The simplest mechanical embedding scheme only
requires a converged NNP that predicts the total energy, forces and
the charges of the ML system in the gas phase but it neglects
polarization of the MM atoms.\cite{toepfer2022fad} In comparison, NNPs
based on electrostatic embedding require additional sampling with MM
atom configurations included as point charges. MD simulation using
ML/MM approaches with electrostatic embedding show great agreement
with MD simulation of respective QM/MM simulation at the same level of
theory as the reference data set.\cite{riniker2021mlmm} The increase
in the quality to describe the impact of the MM solvent on the
properties of the ML system is also demonstrated by accurate
computational reproduction of experimental IR and Raman
spectra.\cite{gastegger2021mlsolv} The most complex polarization
embedding scheme allows the most complete description of the ML system
with the MM environment, but requires more costly reference
computations.\cite{oostenbrink2022burnn} Even a QM/MM model using
polarization embedding is significantly more challenging in terms of
computational effort and implementation than the electrostatic
embedding schemes.\cite{mennucci2020embedding} \\

\section{Applications Based on but Beyond PESs}
Up to this point PESs were used in explicit simulations to determine
experimental observables from dynamics or Monte Carlo
simulations. However, quantum nuclear dynamics or a statistically
significant number of (quasi) classical MD simulations and their
analysis is often a computationally demanding endeavor in itself. It
would be desirable to determine, predict or estimate observables from
only a limited amount of such explicit simulations and devise
rapidly-to-evaluate models that predict with confidence outcomes for
arbitrary input. To set the stage, the full characterization of all
state-to-state cross sections for reactive triatomic systems
A+BC$\rightarrow$AB+C is considered. This problem involves $\sim 10^8$
transitions. Using QCT simulations, convergence of each of the cross
sections requires $\sim 10^5$ independent trajectories to be
run. Hence, for one collision energy $\sim 10^{13}$ QCT simulations
would be required for a full characterization of a reactive triatomic
system. This is neither desirable nor meaningful to do. Hence, despite
the availability of a full-dimensional NN-based or otherwise
represented PES it would be advantageous to reduce the computational
burden of explicitly sampling the PES in this case and the task is to
extract as much information as possible from only a limited number of
simulations.\\

\noindent
The two problems considered further below concern the prediction of
final states or final state distributions for atom+diatom reactions
and predicting thermal rates for bimolecular reactions. Both problems
can, in principle, be solved accurately for carefully chosen systems
which provides the necessary benchmark to extend the range of
applicability of the approaches described below to larger systems.\\

\subsection{Final State Distributions for Atom + Diatom Reactions}
Exhaustive enumeration and characterization of final state
distributions from bimolecular reactions is particularly relevant in
combustion and atmospheric re-entry (hypersonics). The particular
interest is rooted in need to devise more coarse-grained models for
the macroscopic (in space and time) modeling of the chemistry and
physics of reactive flows but based on accurate microscopic
information.\cite{dsmc,MM.hypersonics:2020} For atom+diatom reactions
(A+BC$\rightarrow$AB+C) this involves complete enumeration of all
state-to-state reaction probabilities. As mentioned above, this
problem can - in principle - be addressed by brute-force sampling. But
this is neither practical nor desirable.\\

\noindent
For this reason, ML-based models were devised that allow to either
predict final states or final state distributions from discrete
initial states. From quasiclassical trajectory (QCT) simulations for
the N($^4$S)+NO($^2\Pi$) $\rightarrow$ O($^3$P)+N$_2$(X$^1\Sigma_g^+$)
reaction the state-to-state cross sections $\sigma_{v,j \rightarrow
  v'j'} (E_t)$ as a function of the translational energy $E_t$ were
explicitly determined for 1232 initial ro-vibrational states $(v,j)$
which amounted to $\sim 10^8$ QCT trajectories in total. This compares
with an estimated $10^{15}$ QCT trajectories required for brute-force
sampling of the problem. This information was used as input to train a
NN together with features such as the internal energy, the vibrational
and rotation energy of the diatoms, or the turning points of the
diatoms.\cite{MM.sts:2019} The resulting state-to-state (STS) model is
capable of predicting the cross section for a final state given an
initial collision energy, the vibrational state $v$ of the diatom and
its rotational quantum number $j$. More recently, the approach was
extended to predict entire final state distributions from discrete
initial conditions, which led to the state-to-distribution (STD)
model.\cite{MM.std:2022} Finally, it is also possible to devise
distribution-to-distribution (DTD) models.\cite{MM.dtd:2020}\\

\begin{figure}[b!]
\centering
\includegraphics[width=0.75\textwidth]{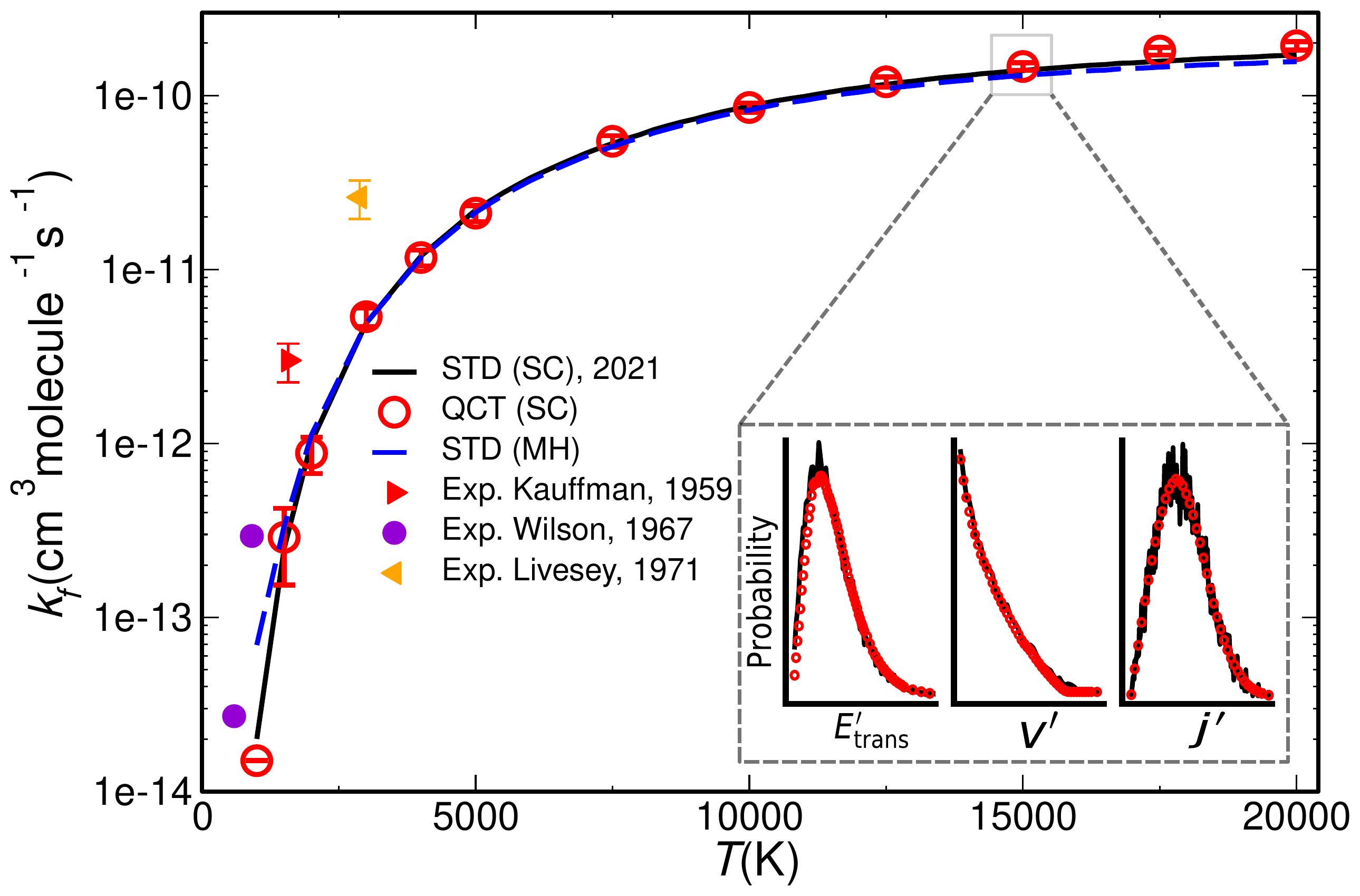}
\caption{The thermal forward rate $k_{f}$ calculated from QCT (open
  red circle) and STD model (solid black line) for the $^{4}$A' state
  of the N($^{4}$S) + O$_{2}$(X$^{3}$ $\Sigma_{g}^{-}$ ) $\rightarrow$
  NO(X$^{2}\Pi$) + O($^{3}$P) reaction between 1000 and 20000
  K. Experimental total forward reaction rate $k_{\rm f}$ (including
  contributions from the doublet and the quartet states) are also
  shown for comparison: (red triangle)\cite{Kaufman:1959}, (orange
  triangle)\cite{Livesey:1971} and (magenta
  circle)\cite{wilson1967rate}. A comparison is made between QCT and
  STD model based on model Hamiltonian $E_{trans}$ (dash blue line)
  for the predicted distributions in the bottom right corner (inset).
  The evaluation is made at $T=15000$ K with QCT and STD evaluations
  marked as black and red solid lines respectively. Figure courtesy
  J. C. San Vicente Veliz.}
\label{fig:std}
\end{figure}

\noindent
The prediction quality of STS, DTD, and STD models is universally high
and reaches a correlation coefficient $R^2 \sim 0.98$ or better
between predicted and QCT-calculated reference data. From these models
it is also possible to determine thermal rates as done for the
N($^4$S)+O$_{2}$(X$^{3}$ $\Sigma_{g}^{-}$ ) reaction shown in
Figure~\ref{fig:std} and further examples are given below.  Comparison
with rates directly determined from QCT simulations - which themselves
are in good agreement with
experiments\cite{MM.cno:2018,MM.no2:2020,MM.co2:2021} - shows that the
trained NNs reach accuracies better than 99 \% over a wide temperature
range $(1000 \leq T \leq 20000)$ K. Thus, ML-based models based on
limited input data from direct simulations on high-quality,
full-dimensional PESs are a computationally efficient and accurate
substitute for explicit, brute-force evaluations of the relevant
properties.\\

\subsection{Predicting Thermal Rates}
Determining thermal rates is one of the major goals of computational
chemistry. Carrying out such a calculation in full dimensionality,
based on an accurate PES and including nuclear quantum effects is a
serious computational undertaking. An accurate rate requires treating
the electronic structure, representing the underlying PES, and running
the (quantum) dynamics simulations at the highest possible levels and
has only been done for a few selected systems. Hence, it is of great
interest to develop models that can predict thermal rates based on
alternative approaches.\\

\noindent
One such effort was based on a library of $\sim 40$ bimolecular
reactions for which $T-$dependent rates from transition state theory
(TST), the Eckart correction to TST, and a set of tabulated ``accurate
rates'' from two-dimensional calculations at 8 temperatures were
available.\cite{houston:2019} These calculations required a
represented PES for carrying out the necessary dynamics
simulations. The data collected was used to learn a correction to the
product of the TST-rate and the Eckart correction by using Gaussian
process regression. Reactions considered included the Cl+HCl H-atom
exchange reaction (in 1d and 3d), the H$_2$+OH$\rightarrow$H+H$_2$O
and for O+CH$_4$ $\rightarrow$ OH+CH$_3$ which was investigated more
in-depth in a separate study.\cite{houston:2022} The results for
reactions not used in the learning procedure indicate that it is
possible to obtain thermal rates close to those from explicit quantum
simulations or trajectory-based quantum calculations (ring polymer
MD).\cite{craig:2005}\\

\subsection{Other Applications}
In one recent application a mapping between local water cluster
arrangement and the frequency of an embedded solute as the
spectroscopic probe was used to predict water anharmonic stretch
vibrations.\cite{cho2021vib} Although this application is not
dependent on and does not require a full-dimensional NN-based PES it
illustrates the potential uses of a mapping between structure and
spectroscopy that can be exploited in the future. Another area which
links intermolecular interactions, structural dynamics and
spectroscopy are ionic and eutectic liquids (ILs and ELs). A strong
case for combining rigorous MD simulations with accurate, ML-based FFs
for property prediction has been made for ionic
liquids.\cite{welton2021ilprops} ILs and ELs are characterized by
strong interactions that probe the short-range part of electrostatics
due to the chemical composition of the systems which consists of a
high density of positively and negatively charged building blocks. For
ELs a recent combination of MD simulations, two-dimensional infrared
and terahertz spectroscopy was able to elucidate the microscopic
structure of the liquid depending on the degree of hydration without,
however, using a ML-based FF.\cite{MM.des:2022} Further improved
agreement between simulations and experiments than that reported can
be expected from refined intermolecular interactions.\\

\section{Challenges}
This section discusses several challenges the field of NN-based PESs
faces. Some of the points discussed may also apply to other ML-based
techniques more broadly in other branches of chemistry. As a very general opening point
it is noted that one of the challenges in statistical approaches is to
extract as much consolidated information, potentially including an
error estimate on the prediction, from a statistical model from as
little information possible. This point concerns very broadly the
aspect of ``data efficiency''.\\
  
\subsection{Data Management and Availability}
Given the tremendous computational cost and effort needed for
generating robust and high quality reference data sets for PES
fitting, data management and availability is a fundamental focus. Yet,
the raw \textit{ab initio} data (nuclear geometries, energies (and
gradients)) of a published PES is often not publicly available,
incomplete or lacks key information such as a precisely specified
level of theory or the employed quantum chemical software. This could
be avoided by publishing exemplary input files alongside the
\textit{ab initio} results. Some of the most popular data sets used
for benchmarking NN potentials contain only equilibrium geometries and
corresponding energies from different levels of theory and are used to
benchmark ML methods. These include the QM7\cite{rupp2012fast},
QM7b\cite{montavon2013machine}, QM9\cite{ramakrishnan2014quantum}, and
ANI-1ccx\cite{smith2020ani} databases. Databases that contain energy
and gradients for equilibrium and distorted structures for different
molecules include ANI-1\cite{smith2017anidata}, the refined
ANI-1x\cite{smith2020ani} and QM7-X.\cite{hoja2021qm7} A popular data
set that provides energies and gradients for configurations visited in
MD simulations is the MD-17 dataset\cite{chmiela2017machine,chmiela2018towards} which is generated
from \textit{ab initio} MD.\\

\noindent
On a cautionary note regarding publicly available datasets, it was
reported that PESs resulting from the MD-17 data are likely to feature
holes in high-energy regions which are visited for example in DMC
simulations.\cite{qu2021multimode,bowman:qm22} Databases such as
ANI-1\cite{smith2017anidata} which uses normal mode sampling for
multiple species also can generate problems. Recently, it was found
that redundancies in databases can compromise the prediction quality
of NN models exploring chemical space.\cite{vazquezsalazar2021} For
training NNPs, the influence of the distribution of the reference
points on the quality of the PES is an open question. Recent efforts
in providing data sets for rigorous and global PES gave rise to the
VIB5\cite{zhang2022vib5} and QM-22\cite{bowman:qm22} databases that
include energies (and gradients) for different molecules calculated at
various levels of theory.\\

\noindent
An often overlooked step in generating databases is the prepossessing
step. It is advisable that the generated data contains as little
redundancies as possible by removing correlated states to reduce the
number of \textit{ab initio} calculations and training time. Therefore,
the generated database can, e.g., be analyzed beforehand by
unsupervised machine learning methods which have been successfully
applied to evaluate MD
trajectories.\cite{ceriotti2019unsupervised,glielmo2021unsupervised}
It is also important to consider that the generation of data must be
application driven because the properties of interest will determine
the amount of data required and should guide the selection of the
sampling method. Data generation for NN-based PESs should be
considered an iterative process in which it is best to start from a
representative and 'clean' data set that will be enriched based on the
problem at hand as was recently done for tunnelling splittings in
malonaldehyde.\cite{kaeser2022tlrpi} \\

\noindent
Finally, ML models are starting to face some of the same difficulties
that the molecular simulation community has been dealing
with.\cite{abraham2019sharing} This includes the lack of standard file
formats, shortage of tools for file sharing, absence of methods to
ensure the quality of the generated databases, etc. Hence, it is worth
mentioning that the young ML community has the unique opportunity to
propose solutions to these obstacles before they become unbearable. In
this regard, the FAIR principle\cite{wilkinson2016fair} (Findable,
Accessible, Interoperable and Reproducible/Reusable) must be taken
into account. In this regard, some authors have proposed general rules
for the application of ML in chemistry\cite{artrith2021best} and in
particular to PESs for small molecules.\cite{li2022data} Specifically,
Li and Liu\cite{li2022data} proposed a checklist for reporting PESs of
high-quality. As a complement to this, we propose some suggestions for
providing data sets underlying molecular PESs.  Data sets should:
\begin{itemize}
    \item provide sample input and output files for the quantum
      chemical software.
    \item have an easy and understandable format.
    \item have a consolidated structure.
    \item contain raw data (at least nuclear geometries, energies (and
      forces)) with clearly defined units, level of theory, employed
      quantum chemical software.
    \item have a clear description of HOW the geometries were
      generated.
    \item if possible, provide information whether the PES was
      developed for a particular purpose/application and whether there
      a known limitations.
    \item be extensible.
\end{itemize}

\subsection{Interpretability}
An important ingredient for extending NN methods is the degree and
conﬁdence with which a human can understand the relationship between
cause (starting database and model) and effect (result or observation,
applying the model to a new
task).\cite{du2019techniques,samek2019towards} This process has also
been called “interpretability”, and it can be used to assess the
relationships learned by the model or contained in the data used for
training.\cite{murdoch2019definitions,dybowski2020interpretable}
However, for complex models like NNs the relationship between input
and output is not clear as a consequence of the non-linearity and
parametric complexity of the models.\cite{keith2021combining}
Therefore, it is not evident if the model is deriving the correct
physics of the system from the provided data or whether it is only
learning artefacts of the data which limits it's application to narrow
settings in what is known as the "clever Hans"
predictor.\cite{lapuschkin2019unmasking} Only a few efforts have been
made to derive techniques that can relate the contribution of
different structural components (atom, bond type) to the predicted
quantity (energy or dipole moment).\cite{schnake2021higher} \\

\noindent
Despite its importance and need, interpretability is still not a main
topic in developing NNPs. A reason for this might be that the use of
conventional techniques is not possible because of the continuous
nature of the properties studied in chemistry.\cite{schutt2019quantum}
However, general guidelines have been
proposed.\cite{letzgus2022toward} By definition, interpretability is
the missing link between the data used for training and the prediction
obtained by the NNP. A better understanding of the inner processes of
NNs will help to better understand the amount of data required to
obtain reliable predictions, understand the \textit{completeness} of
the descriptor, and maybe even some new physical interactions. On the
contrary, the largest risk that the lack of interpretability presents
is that users employ models as 'black box' therefore without knowing
the limitations of the model and possibly obtaining good results for
the wrong reason(s).

\subsection{Generation of Robust Initial Models}
A NNP is only as good as the data it is trained on. As a consequence,
if low-quality data is used the resulting model will under-perform.
This is the principle of 'Garbage in-Garbage Out" which can be traced
back to Charles Babbage.\cite{babbage_2011} The NNP fitting is usually
an iterative process starting from an initial reference data set. This
data set ideally covers the full configurational space of the chemical
system at hand with as few points as possible (note that, in
principle, the number of points on a PES as well as in chemical space
is $\infty$). While an exhaustive sampling of a PES might be possible
for systems with up to 3 atoms (e.g. by choosing configurations on a
regular grid), this becomes impossible for larger
systems. Consequently, the initial sampling relies on (partly random)
methods including MD or normal mode sampling (see Section~\ref{sec4})
that all suffer from distinct weaknesses/disadvantages such as
correlated structures or insufficient coverage. These weaknesses lead
to additional training time, evaluations, \textit{ab initio}
calculations and ultimately to a slower and more expensive convergence
of the iterative NNP fitting procedure.\\

\noindent
Thus, the generation of data for PESs requires improved methods of
(initial) sampling that can warrant sufficient coverage of the PES for
a desired application with as few points as possible.  An interesting
prospect for the generation of PES reference data concerns spreading
the data according to the ``correct'' distribution for different
degrees of freedom resulting from methods like
Boltzmann\cite{Noid2013,reith2003deriving} or Monte Carlo
inversion\cite{lyubartsev1995calculation} and opens the possibility of
deriving interactions from experiments.\cite{chang2022machine} Other
solutions might come from the application of information theory to
ensure a number of samples with the maximum amount of
information. Alternatively, the use of similarity measures between the
initial structures before the actual running of \textit{ab initio}
calculations can be a tool to obtain representative structures of the
PES. However, the problem of how to best choose initial structures for
NNP generation is still open.\\

\noindent
On the other hand, the processing of information by the model can be
enhanced to facilitate the convergence of the model, make it more data
efficient and reduce the dependency on the initial points. This has
been explored for equivariant NNs which complement the description of
the interactions in the message step of MPNNs (See Section
\ref{subsec:arch}). Equivariant NNs have been proven to be very data
efficient by obtaining an accuracy comparable to the best NNPs using
only a fraction of the data that other methods
require.\cite{batzner2022nequip} As a complement to this strategy, it
is possible to obtain data efficient models by including more
physics-based information which has been proved to perform better than
regular approaches for kernel methods.\cite{low2022inclusion}\\

\subsection{Reliable Active Learning and Uncertainty Quantification}
A complete exploration of a PES is a challenging task that most likely
can not be done in a single step and depends heavily on the
application. Therefore, the improvement of PESs is an active topic of
research. Algorithms for systematically improving a training dataset
are known as "active learning" techniques. Active learning is closely
related to uncertainty quantification of the predictions, which by
itself is an active area of research. For NNPs, the most common
technique for obtaining the uncertainty is the training of ensembles
of NNs which are then averaged for the prediction of identical
points. This procedure has a high computationally price because it
requires the training and evaluation of several NNPs. As mentioned
before, ensemble methods present a clear drawback because their
estimated uncertainty can only quantitatively relate to the observed
error.\cite{Kahle2022}\\

\noindent
Other methods of uncertainty quantification like Bayesian NNs, which
impose a prior distribution to each of the parameters of a NNP are
computationally too expensive for practical
use.\cite{gawlikowski2021survey} However, Gaussian processes are a
limiting case of Bayesian NNs\cite{krems2019bayesian}, which have been
extensively used and applied for the refinement of PESs by means of
UQ.\cite{cui2016efficient,vieira2017rate} Therefore, a combination of
NNPs and Gaussian Process Regression is a promising avenue for UQ in
NNPs. Another approach for UQ is single network deterministic
methods\cite{amini2020deep,malinin2020regression} which make
assumptions about the distribution of the data. These methods appear
to be a promising alternative to the mentioned problems by obtaining
the uncertainty by training and evaluating a single model (See Figure
\ref{fig:compound}C). However, it should be noted that single network
models are strongly influenced by the initial assumptions and it is
necessary to calibrate the model beforehand. The need for adjustments
is not an exclusive problem of single network models.  All the
previously described methods require a step of calibration in order to
assure that the predicted uncertainties can be related to the
observed error. Finally, it should be mentioned that active learning
techniques without uncertainty quantification have not been
tested.\cite{settles2012active}\\

\subsection{Extrapolation outside the Training Set Covered}
One of the major drawbacks of NNs is their limited capability to
extrapolate in general beyond the training
data.\cite{haley1992extrapolation} For the case of NNPs this means
that evaluating energies and forces for structures not covered in the
training/validation are likely to lead to a severe breakdown of the
model. This weakness stems from the fact that the functional form
lacks a physical basis and is a pure mathematical fitting
procedure.\cite{behler2011neural} This is different for methods such
as reproducing kernels (RKHS) and PIPs. RKHS allows to choose kernel
functions to follow the physics of the long-range part of the
intermolecular
interactions.\cite{ho1996general,MM.heh2:1999,soldan2000long,unke2017toolkit,MM.heh2:2019}
PIPs make use of Morse variables (i.e. internuclear distances are
usually transformed to Morse variables) which decay to zero for large
distances giving the PES fit a qualitatively correct asymptotic
behaviour\cite{qu2018permutationally}.  However, to obtain the correct
long-range behaviour, PIP PESs often employ switching
functions.\cite{xie2005ab,bowman2009pip} The inability of
extrapolation for NNPs is often revealed at early stages of the NNP
generation and can, e.g., be expressed by unphysical short interatomic
distances or by a partial or entire fragmentation of the
system.\cite{behler2015constructing} Thus, a possible route for
improvement is to include explicit physical knowledge, e.g., on the
long-range electrostatic
interactions\cite{behler2011hdnnp3,yao2018tensormol,unke2019physnet},
dispersion corrections\cite{unke2019physnet}, or on nuclear
repulsion\cite{unke2021spookynet}. Such extensions are likely to allow
extrapolation beyond the training data. Besides the extrapolation in
configurational space, the extrapolation and transferability across
chemical space is of concern.\\

\subsection{Enhancing PESs to Higher Levels of Theory}
Transfer learning and $\Delta$-ML is a comparatively new concept for
theoretical chemistry and solid evaluations are needed. One of the
questions that arises is how to validate the quality of a TL-PES if
single point calculations become increasingly expensive. In other
words, if the effort to carry out one single point \textit{ab initio}
calculation for the HL model required for TL becomes too large, it is
preferable to keep this data in the training set for TL instead of
using it for testing. This certainly gives rise to the question as to
how to probe and validate the NNP for regions that lie outside of the
TL data set. One possible strategy to test the improvement of the HL
PES with respect to the LL PES is to calculate an observable, compare
it to experiment and check for a convergence towards the experiment,
as was done by some of us for the determination of tunneling
splittings.\cite{kaeser2022tlrpi}\\

\noindent
Another open question is what the lowest possible level of theory for
the LL-model is which still allows reliable TL to a HL-model. The
answer to this question will depend on the system and application
considered. Ideally, Hartree-Fock calculations would be a suitable
surrogate model for TL to CCSD(T) levels of theory, but this needs to
be explored for specific systems.\cite{DeltaPaper2015}\\

\noindent
Finally, since the computational cost of the quantum chemical
calculations can be appreciable, again the judicious selection of
molecular structures for which HL calculations are carried out for TL
is crucial. While no simple answer to this question exists as of now,
the structures are usually carefully chosen with human
intervention. Alternatively, it is conceivable that an approach
similar to on-the-fly ML\cite{csanyi2004learn,gastegger2017machine}
(``on-the-fly TL'') could be used to select data points to include in
the TL data set.\\

\subsection{Other Challenges}
Finally, a number of other challenges are briefly summarized. With the
ever increasing quality of NNPs a better understanding of the
relationship between the accuracy of a NNPs based on reference data
for a given quality of the electronic structure and the observables
determined from simulations using this PES is required. Ultimately,
this requires a direct comparison with experiment. This raises the
question whether it is possible to determine the underlying PES from
inverting the relationship between observables and interaction
potential, e.g. by using Invertible NNs.
\cite{ardizzone2018analyzing,arridge2019solving,kothari2021trumpets}
Such an inversion has been done successfully for low-dimensional
systems. The Rydberg-Klein-Rees
(RKR)\cite{rydberg1932graphische,klein1932berechnung,rees1947calculation}
and Rotational RKR (RRKR)\cite{nesbitt1993rotational} procedures are
examples for this. However, for high-dimensional systems, this is a
formidable task and will require a large number of high-quality
data. With respect to the quality of the trained models, more informative
statistical measures should be developed because those used at present
often hide poor performance in individual structures.\\

\noindent
Another challenge ahead is the seamless integration of NNPs - or ML
models in general - into standard MD simulation packages while not
compromising their computational efficiency. Further improvements of
NN-based interaction potentials can be expected from using
physics-informed NNs.\cite{raissi2019physics,wright2022deep} Another
possibility is to explore the combination of NN-based representations
at short range with physics-based long range models based on
multipolar and/or polarizable models.\\

\noindent
Technically, the question arises how complete descriptors need to be
for a comprehensive and accurate representation of the intermolecular
interactions. I.e. what is a meaningful balance between the size of
the descriptor(s) and the accuracy of the final model?  Additionally,
recent advancement in quantum computing technologies provides
opportunities to further reduce the computational cost for generating,
training and applying NNPs.\cite{tao2022exploring} Still, whether and
how these developments will impact how NNPs evolve and are being used
is an open question. \\

\noindent
On the more societal side, it is noted that constructing a
full-dimensional PES for one given molecule is often a computational
investment that requires appreciable resources. Hence, the
environmental impact of this should be considered as
well.\cite{portegies2020ecological,lannelongue2021green,grealey2022carbon}
Generally, all ML-based PESs require the {\it ab initio} computation
of information (energies, forces, or both) for thousands of nuclear
geometries followed by the training of a model which incurs
appreciable environmental cost\footnote{An interesting tool to check
  the $\mathrm{CO_{2}}$ production of your algorithms can be found at:
  http://www.green-algorithms.org/}.\\

\section{Conclusion}
The field of NNPs has reached a considerable degree of maturity in
conceiving PESs that can be used in concrete applications, be it
within the exploration of individual structures or in dynamics-based
studies. Also due to the tremendous progress in efficiency of
electronic structure calculations, it is now possible to determine
full-dimensional - not necessarily ``global'' - potential energy
surfaces for medium-sized molecules at levels of theory that allow
direct comparison and in some cases even prediction of experimental
observables. This, combined with techniques such as transfer learning
holds promise to design yet improved PESs.\\

\noindent
On the other hand, a rather unexplored facet of NNPs concerns
questions about the interpretation of the underlying NN from a
chemical perspective, aspects relating to the optimal distribution of
reference points including minimizing the number of such calculations,
or transferring PESs from one chemical system to a related species
without recomputing all reference information afresh. Solutions to
these questions will considerably increase the efficiency for
conceiving and evaluating NNPs, and improve the prospects for
generalizing trained models to broader chemistries and
applications. \\

\noindent
The present contribution aims at consolidating the available technical
approaches, their use in constructing PESs and their application in
concrete molecular simulations. It is hoped that this will provide a
basis for further development because the prospects of NNPs are bright
and the future for them is open.

\begin{acknowledgement}
This work was supported by the University of Basel, the Swiss National
Science Foundation through grants 200021-117810, 200020-188724, the
NCCR MUST, the Air Force Office for Scientific Research (AFOSR), and
the European Union's Horizon 2020 research and innovation program
under the Marie Sk{\l}odowska-Curie grant agreement No 801459 -
FP-RESOMUS. The authors acknowledge the help of Juan Carlos San
Vicente Veliz for the preparation of Figure \ref{fig:std}. \\
\end{acknowledgement}

\clearpage
\bibliography{references}

\end{document}

%% file: tikz_figs/nn.tex
\begin{tikzpicture}[shorten >=1pt]
		\node[node in](x0) at (0,3.5){};
		\node[node in](x1) at (0,2){};
		\node at (0,1){\vdots};
		\node[node in](xd) at (0,0){};
 
		\node[node hidden](h10) at (3,4){};
		\node[node hidden](h11) at (3,2.5){};
		\node at (3,1.5){\vdots};
		\node[node hidden](h1m) at (3,-0.5){};
 
		\node(h22) at (5,0){};
		\node(h21) at (5,2){};
		\node(h20) at (5,4){};
		
		\node(d3) at (6,0){$\ldots$};
		\node(d2) at (6,2){$\ldots$};
		\node(d1) at (6,4){$\ldots$};
 
		\node(hL12) at (7,0){};
		\node(hL11) at (7,2){};
		\node(hL10) at (7,4){};
		
		\node[node hidden](hL0) at (9,4){};
		\node[node hidden](hL1) at (9,2.5){};
		\node at (9,1.5){\vdots};
		\node[node hidden](hLm) at (9,-0.5){};
 
		\node[node out](y1) at (12,3.5){};
		\node[node out](y2) at (12,2){};
		\node at (12,1){\vdots};	
		\node[node out](yc) at (12,0){};
        \draw (x0) -- (h10);
		\draw (x0) -- (h11);
		\draw (x0) -- (h1m);
        
        \draw (x1) -- (h10);
		\draw (x1) -- (h11);
		\draw (x1) -- (h1m);
 
        \draw (xd) -- (h10);
		\draw (xd) -- (h11);
		\draw (xd) -- (h1m);

		\draw (hL0) -- (y1);
		\draw (hL0) -- (yc);
		\draw (hL0) -- (y2);
 
		\draw (hL1) -- (y1);
		\draw (hL1) -- (yc);
		\draw (hL1) -- (y2);
 
		\draw (hLm) -- (y1);
		\draw (hLm) -- (y2);
		\draw (hLm) -- (yc);
 

        \draw[path fading=east] (h10) -- (h20);
		\draw[path fading=east] (h10) -- (h21);
		\draw[path fading=east] (h10) -- (h22);
		
		\draw[path fading=east] (h11) -- (h20);
		\draw[path fading=east] (h11) -- (h21);
		\draw[path fading=east] (h11) -- (h22);
		
		\draw[path fading=east] (h1m) -- (h20);
		\draw[path fading=east] (h1m) -- (h21);
		\draw[path fading=east] (h1m) -- (h22);
		
		\draw[path fading=west] (hL10) -- (hL0);
		\draw[path fading=west] (hL11) -- (hL0);
		\draw[path fading=west] (hL12) -- (hL0);
		
		\draw[path fading=west] (hL10) -- (hL1);
		\draw[path fading=west] (hL11) -- (hL1);
		\draw[path fading=west] (hL12) -- (hL1);
		
		\draw[path fading=west] (hL10) -- (hLm);
		\draw[path fading=west] (hL11) -- (hLm);
		\draw[path fading=west] (hL12) -- (hLm);

        \node[above=1,align=center,mygreen!60!black] at (0,3) {input\\[-0.2em]layer};
        \node[above=2,align=center,myblue!60!black] at (6,3) {hidden layer};
        \node[above=1,align=center,myred!60!black] at (12,3) {output\\[-0.2em]layer};
        
        \node[above=1,align=center] at (-1,4) {\textbf{\Large A}};
        
	\end{tikzpicture}

%% file: tikz_figs/neuron.tex
\begin{tikzpicture}[
squa/.style={
  draw,
  inner sep=4pt,
  font=\Large
},
start chain=2,node distance=10mm
]

\node[on chain=2,node in] 
  (x2) {$x_2$};
\node[on chain=2] (w2)
  {$w_2$};
  
\node[on chain=2,node hidden] (sigma) 
  {$\displaystyle\Sigma$};

\node[on chain=2,squa,label=above:{\parbox{4cm}{\centering Activation \\ function}}](af)   
  {$f$};
 
\node[on chain=2,label=above:Output,node out] (y)
  {$y$};
  
\begin{scope}[start chain=1]
\node[on chain=1,node in] at (0,1cm) 
  (x1) {$x_1$};
\node[on chain=1] 
  (w1) {$w_1$};
\end{scope}

\begin{scope}[start chain=3]
\node(dots) at (0,-1cm) (x3){\vdots} ;
\node[on chain=3] at (1.3,-1cm) {\vdots};
\end{scope}

\begin{scope}[start chain=4]
\node[on chain=4,node in,label=below:Inputs] at (0,-2cm) 
  (x4) {$x_n$};
\node[on chain=4,label=below:Weights] 
  (w4) {$w_n$};
\end{scope}

\node[label=above:\parbox{2cm}{\centering Bias \\ $b$}] at (sigma|-w1) (b) {};

\draw (x1) -- (w1);
\draw (x2) -- (w2);
\draw (x4) -- (w4);
\draw (w1) -- (sigma);
\draw (w4) -- (sigma);
\draw (w2) -- (sigma);
\draw (b) -- (sigma);
\draw (sigma) -- (af);
\draw (af) --(y);

\node[align=center] at (0,3) {\textbf{\Large B}};
\node[align=center] at (8,3) {\textbf{\Large C}};

\end{tikzpicture}

%% file: tikz_figs/mpnn.tex
  \begin{tikzpicture}[
    operator/.style={
      circle,
      draw=black,
      fill=none,
      font=\small,
      inner sep=1,
      text=black,
      text opacity=1.0},
    mvec/.style={
      draw=black,
      fill=none,
      text=black,
      text opacity=1.0},
    state_h/.style={
      circle,
      fill opacity=0.5,
      draw=none,
      text=black,
      text opacity=1.0,
      minimum size=1.8cm,
      inner sep=0},
    block/.style={
      rectangle,
      draw=black,
      thick,
      align=center,
      rounded corners,
      minimum height=2em}]
  
  \node[] at (0,9) {$\boldsymbol i$};
  \node[] at (2.5,9) {$\boldsymbol 1$};
  \node[] at (5.5,9) {$\boldsymbol 2$};
  \node[] at (8.5,9) {$\boldsymbol 3$};
  
  \node[] at (0,8) {$\boldsymbol h_i^{t=0}$};
  \node (h01) at (3,8) 
    [state_h, ball color={rgb,1:red,1;green,0;blue,0},] 
    {$\left( 1,0,0 \right)$};
  \node (h02) at (6,8) 
    [state_h, ball color={rgb,1:red,0;green,1;blue,0},] 
    {$\left( 0,1,0 \right)$};
  \node (h03) at (9,8) 
    [state_h, ball color={rgb,1:red,0;green,0;blue,1},] 
    {$\left( 0,0,1 \right)$};
  
  \begin{pgfonlayer}{background}
    \draw[gray, line width=8] (h01) -- (h02) -- (h03);
  \end{pgfonlayer}
  
  \draw[thick,<->]  (h01) -- (h02) node[pos=0.5,above] {$R$};
  
  \draw[thick,<->]  (h01) -- (6.8,10.0) node[pos=0.3,,above] {$R_c$};
  \draw[dotted,very thick] (7.5,8.2) arc (5:35:3.8);
  
  \node[block, rotate=90] at (-0.5,5.5) {Interaction layer};
  
  \node[] at (0.75,6.5) {$\boldsymbol M_{t=0}$};
  \node (o01) [operator] at (3,6.5) {$+$};
  \node (o02) [operator] at (6,6.5) {$+$};
  \node (o03) [operator] at (9,6.5) {$+$};
  
  \draw[thick]  (h01) -- (o01);
  \draw[thick]  (h02) -- (o01);
  \draw[thick]  (h01) -- (o02);
  \draw[thick]  (h02) -- (o02);
  \draw[line width=4, color={rgb,1:red,0;green,0;blue,1}]  
    (h03) -- (o02);
  \draw[thick]  (h02) -- (o03);
  \draw[thick]  (h03) -- (o03);
  
  \node[] at (0.75,5.5) {$\boldsymbol m_i^{t=0}$};
  \node (m01) [mvec] at (3,5.5) 
    {$\left( 1,1,0 \right)$};
  \node (m02) [mvec] at (6,5.5)
    {$\left( 1,1,{\color{blue} 1} \right)$};
  \node (m03) [mvec] at (9,5.5)
    {$\left( 0,1,1 \right)$};
    
  \draw[thick]  (o01) -- (m01);
  \draw[line width=4, color={rgb,1:red,0;green,0;blue,1}]
    (o02) -- (m02);
  \draw[thick] (o03) -- (m03);
  
  \node[] at (0.75,4.5) {$\boldsymbol U_{t=0}$};
  \node (u01) [operator,font=\scriptsize] at (3,4.5) {${\tiny|}\!+\!{\tiny|}$};  
  \node (u02) [operator,font=\scriptsize] at (6,4.5) {${\tiny|}\!+\!{\tiny|}$};  
  \node (u03) [operator,font=\scriptsize] at (9,4.5) {${\tiny|}\!+\!{\tiny|}$};  

  \draw[thick]  (m01) -- (u01);
  \draw[line width=4, color={rgb,1:red,0;green,0;blue,1}]
    (m02) -- (u02);
  \draw[thick]  (m03) -- (u03);

  \draw[dashed,thick]  (h01) -- (1.75,6.5);
  \draw[dashed,thick]  (h02) -- (4.75,6.5);
  \draw[dashed,thick]  (h03) -- (7.75,6.5);
  \draw[dashed,thick]  (1.75,6.5) -- (1.75,4.5);
  \draw[dashed,thick]  (4.75,6.5) -- (4.75,4.5);
  \draw[dashed,thick]  (7.75,6.5) -- (7.75,4.5);
  \draw[dashed,thick]  (1.75,4.5) -- (u01);
  \draw[dashed,thick]  (4.75,4.5) -- (u02);
  \draw[dashed,thick]  (7.75,4.5) -- (u03);

  \node[] at (0,3) {$\boldsymbol h_i^{t=1}$};
  \node (h11) at (3,3)
    [state_h, font=\small, ball color={rgb,3:red,2;green,1;blue,0}] 
    {$\left( 
      \dfrac{2}{3},
      \dfrac{1}{3},
      0 
    \right)$};
  \node (h12) at (6,3)
    [state_h, font=\small, ball color={rgb,4:red,1;green,2;blue,1}] 
    {$\left( 
      \dfrac{1}{4},
      \dfrac{1}{2},
      \dfrac{1}{4}
    \right)$};
  \node (h13) at (9,3)
    [state_h, font=\small, ball color={rgb,3:red,0;green,1;blue,1}] 
    {$\left( 
      0,
      \dfrac{1}{3},
      \dfrac{2}{3}
    \right)$};

  \begin{pgfonlayer}{background}
    \draw[gray, line width=8] (h11) -- (h12) -- (h13);
  \end{pgfonlayer}

  \draw[thick] (u01) -- (h11);
  \draw[line width=4, color={rgb,1:red,0;green,0;blue,1}]
   (u02) -- (h12);
  \draw[thick] (u03) -- (h13);
  
  \node[block, rotate=90] at (-0.5,0.5) {Interaction layer};
  
  \node[] at (0.75,1.5) {$\boldsymbol M_{t=1}$};
  \node (o11) [operator] at (3,1.5) {$+$};
  \node (o12) [operator] at (6,1.5) {$+$};
  \node (o13) [operator] at (9,1.5) {$+$};
  
  \draw[thick]  (h11) -- (o11);
  \draw[line width=2, color={rgb,1:red,0;green,0;blue,1}]
    (h12) -- (o11);
  \draw[thick]  (h11) -- (o12);
  \draw[thick]  (h12) -- (o12);
  \draw[thick]  (h13) -- (o12);
  \draw[thick]  (h12) -- (o13);
  \draw[thick]  (h13) -- (o13);
  
  \node[] at (0.75,0.5) {$\boldsymbol m_i^{t=1}$};
  \node (m11) [mvec, font=\tiny] at (3,0.5) 
    {$\left( 
      \dfrac{11}{12},
      \dfrac{5}{6},
      {\color{blue} \dfrac{1}{4}}
    \right)$};
  \node (m12) [mvec, font=\tiny] at (6,0.5)
    {$\left( 
      \dfrac{11}{12},
      \dfrac{7}{6},
      \dfrac{11}{12}
    \right)$};
  \node (m13) [mvec, font=\tiny] at (9,0.5)
    {$\left( 
      \dfrac{1}{4},
      \dfrac{5}{6},
      \dfrac{11}{12}
    \right)$};
    
  \draw[line width=2, color={rgb,1:red,0;green,0;blue,1}]
    (o11) -- (m11);
  \draw[thick]  (o12) -- (m12);
  \draw[thick]  (o13) -- (m13);
  
  \node[] at (0.75,-0.5) {$\boldsymbol U_{t=1}$};
  \node (u11) [operator,font=\scriptsize] at (3,-0.5) {${\tiny|}\!+\!{\tiny|}$};
  \node (u12) [operator,font=\scriptsize] at (6,-0.5) {${\tiny|}\!+\!{\tiny|}$};
  \node (u13) [operator,font=\scriptsize] at (9,-0.5) {${\tiny|}\!+\!{\tiny|}$};  

  \draw[line width=2, color={rgb,1:red,0;green,0;blue,1}]
    (m11) -- (u11);
  \draw[thick]  (m12) -- (u12);
  \draw[thick]  (m13) -- (u13);

  \draw[dashed,thick]  (h11) -- (1.75,1.5);
  \draw[dashed,thick]  (h12) -- (4.75,1.5);
  \draw[dashed,thick]  (h13) -- (7.75,1.5);
  \draw[dashed,thick]  (1.75,1.5) -- (1.75,-0.5);
  \draw[dashed,thick]  (4.75,1.5) -- (4.75,-0.5);
  \draw[dashed,thick]  (7.75,1.5) -- (7.75,-0.5);
  \draw[dashed,thick]  (1.75,-0.5) -- (u11);
  \draw[dashed,thick]  (4.75,-0.5) -- (u12);
  \draw[dashed,thick]  (7.75,-0.5) -- (u13);

  \node[] at (0,-2) {$\boldsymbol h_i^{t=2}$};
  \node (h21) at (3,-2)
    [state_h, font=\tiny, ball color={rgb,36:red,19;green,14;blue,3}] 
    {$\left( 
      \dfrac{19}{36},
      \dfrac{7}{18},
      \dfrac{1}{12}
    \right)$};
  \node (h22) at (6,-2)
    [state_h, font=\tiny, ball color={rgb,24:red,7;green,10;blue,7}] 
    {$\left( 
      \dfrac{7}{24},
      \dfrac{5}{12},
      \dfrac{7}{24}
    \right)$};
  \node (h23) at (9,-2)
    [state_h, font=\tiny, ball color={rgb,36:red,3;green,14;blue,19}] 
    {$\left( 
      \dfrac{1}{12},
      \dfrac{7}{18},
      \dfrac{19}{36}
    \right)$};

  \begin{pgfonlayer}{background}
    \draw[gray, line width=8] (h21) -- (h22) -- (h23);
  \end{pgfonlayer}

  \draw[line width=2, color={rgb,1:red,0;green,0;blue,1}]
    (u11) -- (h21);
  \draw[thick] (u12) -- (h22);
  \draw[thick] (u13) -- (h23);
  
  \end{tikzpicture}